\documentclass[aps,prd,preprint,superscriptaddress,showpacs,floatfix,preprintnumbers,nofootinbib,
a4paper]{revtex4-1}
\usepackage{hyperref}
\usepackage{graphicx}
\usepackage{color}
\usepackage{graphics}
\usepackage{subfigure}
\usepackage{amsmath, amssymb, epsfig}
\newcommand{\mathsym}[1]{{}}

\newcommand{\ba}{\begin{array}}
\newcommand{\ea}{\end{array}}
\newcommand{\bal}{\begin{align}}
\newcommand{\eal}{\end{align}}
\newcommand{\be}{\begin{equation}}
\newcommand{\ee}{\end{equation}}
\newcommand{\beqa}{\begin{eqnarray}}
\newcommand{\eeqa}{\end{eqnarray}}
\def\321{$SU(3)\times SU(2)\times U(1)$}

\newcommand{\dms}{\Delta m^2_{\rm sol}}
\newcommand{\Dma}{\Delta m^2_{\rm atm}}
\def\10{SO(10)}

\def\se{\tilde{e}}
\def\smu{\tilde{\mu}}
\def\stau{\tilde{\tau}}

\newcommand{\brg}[2]{{\rm BR}(#1 \rightarrow #2 \gamma)}

\newcommand{\lrb}{ \left( }
\newcommand{\rrb}{ \right) }
\newcommand{\lcb}{ \left\{ }
\newcommand{\rcb}{ \right\} }
\newcommand{\lsb}{ \left[ }
\newcommand{\rsb}{ \right] }

\hypersetup{
 pdfstartview=FitV,
 colorlinks=true, 
 linkcolor=blue, 
 citecolor=red, 
 filecolor=black, 
 urlcolor=black}

\begin{document}
\title{Revisiting lepton flavor violation in supersymmetric type II seesaw}

\author{Debtosh Chowdhury}
\email{debtosh@cts.iisc.ernet.in}
\affiliation{Centre for High Energy Physics, Indian Institute of Science, Bangalore 560 012, India.}
\author{Ketan M. Patel}
\email{ketan@theory.tifr.res.in}
\affiliation{Department of Theoretical Physics, Tata Institute of Fundamental
Research, Mumbai 400 005, India. \vspace*{0.5cm}}

\preprint{TIFR/TH/13-12}
\date{\today}
\vspace*{2cm}

\begin{abstract}
\noindent In view of the recent measurement of reactor mixing angle $\theta_{13}$ and updated limit
on $\brg{\mu}{e}$ by the MEG experiment, we re-examine the charged lepton flavor violations in
a framework of supersymmetric type II seesaw mechanism. Supersymmetric type II seesaw predicts
strong correlation between $\brg{\mu}{e}$ and $\brg{\tau}{\mu}$ mainly in terms of the neutrino
mixing angles. We show that such a correlation can be determined accurately after the measurement of
$\theta_{13}$. We compute different factors which can affect this correlation and show that the
mSUGRA-like scenarios, in which slepton masses are taken to be universal at the high scale, predicts
$3.5\lesssim \brg{\tau}{\mu}/\brg{\mu}{e} \lesssim 30$ for normal hierarchical neutrino masses. Any
experimental indication of deviation from this prediction would rule out the minimal models of
supersymmetric type II seesaw. We show that the current MEG limit puts severe constraints on the
light sparticle spectrum in mSUGRA model if the seesaw scale lies within $10^{13}$-$10^{15}$ GeV. It
is shown that these constraints can be relaxed and relatively light sparticle spectrum can be
obtained in a class of models in which the soft mass of triplet scalar is taken to be non-universal
at the high scale.
\end{abstract}

\maketitle

\section{Introduction}
\label{intro}

The idea of seesaw mechanism \cite{Minkowski:1977sc,*Yanagida:1979as,*GellMann:1980vs,
*Mohapatra:1979ia,Schechter:1980gr,*Mohapatra:1980yp,*Lazarides:1980nt,*Cheng:1980qt,Foot:1988aq,
*Ma:1998dn} is perhaps the most elegant way to account for the tiny
neutrino masses evidenced from various neutrino oscillation experiments in the last two
decades. Tree-level realization of seesaw mechanism requires an extension of the Standard Model
(SM) by either heavy fermion singlets or a heavy scalar triplet or heavy fermion triplets. In
the literature, these versions are famously known as the seesaw mechanisms of type I
\cite{Minkowski:1977sc,*Yanagida:1979as,*GellMann:1980vs,*Mohapatra:1979ia}, II
\cite{Schechter:1980gr,*Mohapatra:1980yp,*Lazarides:1980nt,*Cheng:1980qt} and III
\cite{Foot:1988aq,*Ma:1998dn} respectively. The scale at which these new
fields decouple from the SM is known as the seesaw scale. Present information from the neutrino
oscillation data suggests that such a scale should be in the range $10^{9}$-$10^{15}$ GeV if no
artificial fine tuning is assumed in the Yukawa coupling parameters. Seesaw also generates mixing
among the neutrino flavors and it leads to lepton flavor violating (LFV) effects in the charged
lepton sector, for example,
decays like $l_i \to l_j \gamma$ which are otherwise absent in the SM. Such decays are mediated by
the heavy fields required for tree-level realization of seesaw mechanism and so, at least in
principle, they can shed a light on the exact nature of seesaw mechanism. However such effects
are extremely tiny in the seesaw extensions of SM due to the ultra heavy seesaw scale. As a result,
in these class of theories, one can not distinguish between the various tree-level realizations of
the seesaw mechanism even if they are truly responsible for small neutrino masses.

Low energy supersymmetry (SUSY) can provide insight to the seesaw mechanism by mediating LFV decays
through sparticles. In the minimal models of supersymmetry like the constrained minimal
supersymmetric standard model (CMSSM) or mSUGRA, the SUSY breaking Lagrangian conserves the
lepton flavors since all the soft scalar masses and trilinear $A$-terms are taken to be universal at
the scale of grand unification (GUT). The LFVs are present only in the Yukawa couplings at this
scale in the SUSY conserving superpotential extended suitably to accommodate the seesaw mechanism.
Now, if the seesaw scale is slightly lower than the GUT scale, mixings among the sleptons of
different generation get induced at the seesaw scale through (i) renormalization group evolution (RGE) effects and (ii) lepton flavor violating Yukawa couplings. As a result, slepton mass matrices no longer remain diagonal at the seesaw scale. At low energy, the off-diagonal entries in the slepton mass matrices induce large rate of LFV decays through
one-loop diagrams which involves the gaugino exchange \cite{Lee:1984kr,*Lee:1984tn,*Borzumati:1986qx,*Hall:1985dx,*Gabbiani:1988rb}. 
These effects are extensively studied in the literature in the context of all three variants of the seesaw mechanism
\cite{Hisano:1995nq,*Hisano:1995cp,*Hisano:1998fj,*Masiero:2002jn,*Babu:2002tb,*Masiero:2004js,
*Calibbi:2006nq,*Antusch:2006vw,*Calibbi:2006ne,*reviews1,*herrero,*Joshipura:2009gi,*Biggio:2010me,
*Esteves:2010ff,*Cannoni:2013gq,Calibbi:2012gr,Rossi:2002zb,Joaquim:2006uz,*Joaquim:2006mn,
*Joaquim:2009vp,Arganda:2005ji,*Hirsch:2008gh,*Esteves:2009qr,*Calibbi:2009wk,*Hirsch:2012ti}.

An exact determination of the LFV rates requires complete knowledge of the Yukawa coupling
matrices which violate the lepton flavors. 
In the type I and III seesaws, LFVs occur through the Dirac neutrino Yukawa
couplings with $SU(2)_L$ singlet and triplet fermions respectively. Determination of such couplings
from the physical neutrino parameters is not straight forward using the seesaw formula since it also
requires the knowledge of additional parameters present in the mass matrix of singlet or triplet
fermions. The type II seesaw, on the contrary, invokes only one Yukawa coupling matrix
\cite{Schechter:1980gr,*Mohapatra:1980yp,*Lazarides:1980nt,*Cheng:1980qt} which can completely be
determined from the low energy values of neutrino masses and mixing angles up to an overall factor
and RGE effects \cite{Rossi:2002zb}. 
This property makes type II seesaw framework most predictive among all three seesaw scenarios. 


It was pointed out in \cite{Rossi:2002zb} that due to its predictive power, SUSY version of type II
seesaw can actually predict the ratios of decay rates of different LFV channels in terms of the
neutrino mixing angles and solar and atmospheric mass squared differences. This was further
analyzed in \cite{Joaquim:2006uz,*Joaquim:2006mn,*Joaquim:2009vp} through detailed numerical
analysis. These ratios can provide a powerful probe for distinguishing type II seesaw from the other
versions of seesaw mechanism if LFV is observed in at least one of the decay channels like $l_i \to
l_j \gamma$. However, such correlations depends on the reactor mixing angle $\theta_{13}$ which was
not known till mid 2011. Thanks to the several reactor oscillation experiments
\cite{Abe:2011sj,*Adamson:2011qu,*Abe:2011fz,*An:2012eh,*Ahn:2012nd}, $\theta_{13}$ is now precisely
known. The global fit of neutrino oscillation data gives \cite{GonzalezGarcia:2012sz}
\be \label{theta13}
\sin^2 \theta_{13} = 0.023 \pm 0.0023\ . \ee
We show that the above measurement fixes the ratios like BR($\tau \to \mu \gamma$)/BR($\mu \to
e \gamma$) in the type II seesaw case. Further, we show that such ratios can be ``fudged'' if there
exists a large hierarchy among the sleptons of different generations. We determine such fudge factor
as a function of soft SUSY breaking parameters in mSUGRA-like models and show that one can still
make a robust prediction for such ratios. We also discuss the effects of leptonic CP violation on
the prediction of ratio BR($\tau \to \mu \gamma$)/BR($\mu \to e \gamma$).

Charged lepton flavor violations in SUSY type II seesaw models have been studied in several works 
\cite{Rossi:2002zb,Joaquim:2006uz,*Joaquim:2006mn,*Joaquim:2009vp,Arganda:2005ji,*Hirsch:2008gh,
*Esteves:2009qr,*Calibbi:2009wk,*Hirsch:2012ti}. We
revisit them in the context of some recent progress made in the
experimental searches of new physics. These include the discovery of Higgs like boson at the LHC,
the measurement of $\theta_{13}$, negative results in all the direct searches of SUSY and updated
results on some of the indirect searches in $B$-decays like $B \to X_s \gamma$,  $B_s \to \mu^+
\mu^-$ etc. Further, the MEG collaboration has very recently updated the upper
limit on BR$(\mu \to e \gamma)$ which is now $5.7 \times 10^{-13}$ \cite{Adam:2013mnn}, an order 
of magnitude stronger than the previous limit. The experimental limits on the other charged LFV
decays, as summarized in Table \ref{lfv-exp}, are not so strong compared to BR$(\mu \to e \gamma)$.
Altogether, the above experimental constraints put severe limits on the light sparticle spectrum in
CMSSM/mSUGRA like models as we show in this paper. 
\begin{table}[!ht]
 \begin{center}
 \begin{tabular}{lll}
 \hline  \hline
 LFV Process ~~~~~&Present bound   ~~~~~&Near future sensitivity  \\
 & ~~~~~&of ongoing experiments\\
 \hline
 BR($\mu\rightarrow e \gamma$) & $5.7 \times 10^{-13}$\ \cite{Adam:2013mnn}
& $6 \times 10^{-14}$\ \cite{Baldini:2013ke} \\
 BR($\tau\rightarrow e \gamma$) & $3.3 \times 10^{-8}$\ \cite{Amhis:2012bh} &
$10^{-9}$\ \cite{Brodzicka:2012jm} \\
 BR($\tau\rightarrow \mu \gamma$) & $4.4 \times 10^{-8}$\ \cite{Amhis:2012bh} &
$3 \times 10^{-9}$\ \cite{Brodzicka:2012jm} \\
 BR($\mu\rightarrow e e e$) & $1.0 \times 10^{-12}$\ \cite{Bellgardt:1987du} & $10^{-15}$\
\cite{Blondel:2013ia} \\
 BR($\tau\rightarrow e e e$) & $3.0 \times 10^{-8}$\ \cite{Amhis:2012bh} & $10^{-9}$\
\cite{Brodzicka:2012jm}\\ 
BR($\tau\rightarrow \mu \mu \mu$) & $2.0 \times 10^{-8}$\ \cite{Amhis:2012bh} & $3 \times 10^{-9}$\
\cite{Brodzicka:2012jm} \\
 \hline
 \hline
 \end{tabular}
\end{center}
\caption{Present bounds and future sensitivities expected from the current generation experiments on
the various LFV processes.}
\label{lfv-exp}
\end{table}

It is recently noted in \cite{Calibbi:2012gr} that novel cancellation in the magnitude of charged LFVs can
arise
in the case of type I seesaw models if the soft masses of MSSM Higgs doublets are taken to be
different from the sfermion masses at the GUT scale. We find that similar cancellation can also
arise in type II seesaw if the soft mass of triplet scalar is different from the soft masses of
sfermions and Higgs doublets at the GUT scale. We identify such scenario as the non-universal
triplet model (NUTM) and perform a detailed numerical analysis for it. In this class of models,
the constraints from charged LFVs can be evaded up to the certain extent which opens a room for
relatively light SUSY spectrum.

The paper is organized as the following. We review the SUSY version of type II seesaw mechanism in
the next section. In Section \ref{msugra}, we study in detail the type II seesaw in mSUGRA like
models and present the results obtained from the numerical analysis. The results of similar
analysis conducted for NUTM model are presented in Section \ref{nutm}. Finally, we summarize in
Section \ref{summary}.

\section{Supersymmetric type II seesaw mechanism and lepton flavor violation}
\label{type-II-lfv}
We briefly review here type II seesaw mechanism in order to set the stage for the relevant LFV
studies. In type II seesaw, the neutrino masses arise from their Yukawa interaction with $SU(2)_L$
triplet superfield $T$ which has the hypercharge $Y=2$. A conjugate superfield $T^c \sim (3, -2)$ is
required in supersymmetric versions to cancel the anomalies. The relevant part of the
superpotential can be written as
\cite{Rossi:2002zb,Joaquim:2006uz,*Joaquim:2006mn,*Joaquim:2009vp,Hirsch:2008gh,*Esteves:2009qr,
*Calibbi:2009wk,*Hirsch:2012ti}
\be \label{suppot}
W_T = \frac{1}{\sqrt{2}} \lsb \lrb{{\bf Y}_T}\rrb_{ij} L_i T L_j + \lambda_u H_u T^c H_u + 
\lambda_d H_d T H_d \rsb + M_T T T^c, \ee
where ${\bf Y}_T$ is in general complex symmetric matrix in the generation space and $H_{u,d}$ are
the standard MSSM Higgs doublets with hypercharge $Y=\pm 1$. Note that the full
superpotential explicitly breaks the lepton number (LN) conservation irrespective of any choice of
LNs for $T$ and $T^c$. At the scale much below $M_T$, Eq. (\ref{suppot}) generates the masses for
neutrinos after the electroweak symmetry is broken, namely
\be \label{numass}
{{\cal M}_\nu} = \frac{v_u^2 \lambda_u}{M_T} {\bf Y}_T~. \ee
Here, $v_u \equiv \langle H_u \rangle/\sqrt{2} = v \sin \beta$ and $ v = 174$ GeV. Clearly, Eq.
(\ref{numass}) makes possible to determine the only source of lepton flavor violation up to an
overall constant, {\it i.e.} ${\bf Y}_T$ at the high scale, in terms of the data of neutrino masses and
mixing angles extrapolated at the scale $M_T$. As already noted in \cite{Rossi:2002zb}, this 
feature of type II seesaw makes it more predictive scenario for the calculations of LFVs
compared to type I and type III seesaws. The overall constant can be estimated using the scale of
atmospheric squared mass difference. {\it i.e.} $M_T \approx \lambda_u v_u^2/\sqrt{\Dma}$, where
$\Dma\equiv |m_{\nu_3}^2-m_{\nu_1}^2|$. In Eq. (\ref{numass}), the requirement of perturbative ${\bf
Y}_T$ and $\lambda_u$ puts an upper limit on the seesaw scale
\be \label{MT-limit}
M_T \lesssim 6\times 10^{14}\ {\rm GeV}. \ee

An extra pair of triplets $T$ and $T^c$ in the MSSM with mass $M_T$ significantly below the
unification scale $M_{\rm GUT}$ spoils the gauge coupling unification. As it is very well-known,
this problem can be avoided if a pair of full $\mathbf{15} + \overline{\bf 15}$ multiplet of $SU(5)$ 
is added in the spectrum \cite{Rossi:2002zb,Joaquim:2006uz,*Joaquim:2006mn,*Joaquim:2009vp}. In addition to a triplet $T$, the ${\bf 15}$-plet contains two more
multiplets $S\sim(6,1,-4/3)$ and $Z\sim(3,2,1/3)$ which restore the gauge coupling unification if
the masses of all three multiplets are degenerate. Furthermore, along with triplet, the extra
multiplets $S$ and $Z$ also have Yukawa interactions with MSSM matter fields like ${\bf Y}_S D^c S
D^c$ and  ${\bf Y}_Z D^c Z L$ which arise from a common Yukawa term ${\bf Y}_{\bf 15}\ \overline{\bf 5}
\cdot {\bf 15} \cdot  \overline{\bf 5} $ in $SU(5)$. Considering the above $SU(5)$ GUT
scenario as a minimal framework for type II seesaw mechanism consistent with the gauge coupling
unification, we impose $M_T=M_S=M_Z\equiv M_{\bf 15}$ and ${\bf Y}_T = {\bf Y}_S = {\bf Y}_Z \equiv {\bf Y}_{\bf 15}$ at the GUT scale. Note that the above mass equality condition gets destroyed at $M_T$ due
to RG running from $M_{\rm GUT}$ to $M_T$. However it still maintains the gauge coupling unification
to good accuracy since such effects are very small. The entire pair of ${\bf 15}+\overline{\bf 15}$
multiplets get decoupled from the spectrum below $M_T$. 

We now turn our discussion to charged LFVs in the context of above scenario. As already
noted in Section \ref{intro}, flavor violation in charged leptons arises through the mixings
between sleptons of different flavors in SUSY versions of seesaws. Such mixings get induced due
to RG running from $M_{\rm GUT}$ to $M_T$ even if they are absent at $M_{\rm GUT}$ in the models
with universal boundary conditions. The presence of a pair of triplet scalar between $M_T$ and
$M_{\rm GUT}$ generates small off-diagonal elements in the left-handed slepton mass
matrix $\mathbf{m}^2_{\tilde L}$. In the leading logarithmic approximation, solution to the
1-loop RGE equation of $\mathbf{m}^2_{\tilde L}$ can be estimated as \cite{Rossi:2002zb} :
\be \label{Delta1}
\lrb \mathbf{m}^2_{\tilde L} \rrb_{i \neq j} \approx -\frac{3(3m_0^2+A_0^2)}{8 \pi^2}({\bf
Y}_T^\dagger {\bf Y}_T)_{ij}
\log\left( \frac{M_{\rm GUT}}{M_T}\right), \ee
where $m_0$ is the universal soft mass for scalars and $A_0$ is the universal trilinear coupling 
defined at the GUT scale. Using Eq. (\ref{numass}), one can write
\be \label{YT}
{\bf Y}_T^\dagger {\bf Y}_T = \left( \frac{M_T}{v_u^2 \lambda_u}\right)^2 {\bf U}_{\rm MNS}\ {\rm Diag.}
\lrb m_{\nu_1}^2,\ m_{\nu_2}^2,\ m_{\nu_3}^2 \rrb\ {\bf U}_{\rm MNS}^\dagger, \ee
where ${\bf U}_{\rm MNS}$ is Maki-Nakagawa-Sakata lepton mixing matrix which can be parametrized 
by three mixing angles, one Dirac phase and two Majorana phases in the standard parametrization \cite{Beringer:1900zz}. It is trivial to check that Majorana phases do not contribute in
$ \lrb \mathbf{m}^2_{\tilde L} \rrb_{ij}$ at 1-loop level. For the following discussions, we neglect
Dirac CP phase for simplicity. As can be seen from Eq. (\ref{YT}), one can determine ${\bf Y}_T^\dagger {\bf Y}_T$ at $M_{T}$ from the extrapolated values of neutrino masses and mixing parameters, up to an overall constant.


The branching ratio of a general charged LFV decay $l_i\rightarrow l_j\gamma$ ($i\neq j$) can be
estimated as \cite{Borzumati:1986qx}
\be \label{general-lfv}
{\rm BR}(l_i\rightarrow l_j\gamma) \approx \frac{\alpha^3}{G_F^2}
\frac{\left| \delta^{LL}_{ij} \right| ^2}{M_{\rm SUSY}^4} \tan^2\beta\times{\rm BR}(l_i\rightarrow l_j \nu_i
\bar{\nu}_j)\, \ee
where $M_{\rm SUSY}$ is the generic SUSY breaking scale and the flavor violations are parametrized 
as
\be \label{delLL}
\delta^{LL}_{ij} \equiv \frac{\lrb \mathbf{m}^2_{\tilde L} \rrb_{ij}}{\bar{m}_{\tilde{l}_i} 
\bar{m}_{\tilde{l}_j}}\,
\ee
where $\bar{m}_{\tilde{l}}\equiv \sqrt{m_{\tilde{l}_1} m_{\tilde{l}_2}}$ is the geometric mean of 
the masses of sleptons involved in the process. Before we present detailed numerical analysis of LFVs in
mSUGRA and extended models, let us briefly discuss some qualitative features that emerge from Eq.
(\ref{general-lfv}). 

\subsection{The seesaw scale dependency of the branching ratios}
\label{sect-2a}
One would typically expect from Eq. (\ref{Delta1}) that the relatively low triplet scale $M_T$
enhances the flavor violations through large running effects. However note that in Eq. (\ref{YT}) the small $M_T$
also decreases ${\bf Y}_T$ unless $\lambda_u$ is tuned accordingly. It can be seen from Eqs.
(\ref{Delta1}, \ref{YT}, \ref{general-lfv}, \ref{delLL}) that 
\be \label{mt-dependence}
{\rm BR}(l_i\rightarrow l_j\gamma) \propto \left|\lrb{\bf Y}_T^\dagger {\bf Y}_T\rrb_{ij}\right|^2 
\propto \left( \frac{M_T m_{\nu_3}}{v_u^2 \lambda_u} \right)^4
\ee
at a given SUSY scale. If $\lambda_u $ is taken to be of ${\cal O}$(1), the light seesaw scale
leads to the smaller values of ${\bf Y}_T$ as required to fit the light neutrino masses. This
significantly suppresses the rates of charged LFV processes.

\subsection{Correlations among the branching ratios of different decay channels}
\label{sect-2b}
One finds from Eqs. (\ref{general-lfv}, \ref{delLL})
\be \label{ratio}
\frac{{\rm BR}(\tau \rightarrow \mu \gamma)}{{\rm BR}(\mu \rightarrow e \gamma)} \approx 
\left\lvert\frac{(\mathbf{m}^2_{\tilde L})_{\tau \mu}}{(\mathbf{m}^2_{\tilde L})_{\mu
e}}\right\rvert^2\
\frac{\bar{m}_{\se}^2}{\bar{m}_{\stau}^2}\ \frac{{\rm BR}(\tau \rightarrow \mu \nu_\tau
\bar{\nu}_\mu)}{{\rm BR}(\mu \rightarrow e \nu_\mu \bar{\nu}_e)}\ . \ee 
The ratio of lepton flavor conserving branching ratios, namely  ${\rm BR}(\tau \rightarrow \mu
\nu_\tau \bar{\nu}_\mu)/{\rm BR}(\mu \rightarrow e \nu_\mu \bar{\nu}_e)$, is  0.18. Further,
using  Eqs. (\ref{Delta1}, \ref{YT}), one obtains for the normal hierarchy
in neutrino masses ({\it i.e.} $m_{\nu_1}\approx 0$, $m_{\nu_2}\approx\sqrt{\dms}$ and
$m_{\nu_3}\approx\sqrt{\Dma}$) and for non-zero $\theta_{13}$
\be \label{ratio-ml}
\left\lvert\frac{(\mathbf{m}^2_{\tilde L})_{\tau \mu}}{(\mathbf{m}^2_{\tilde L})_{\mu e}}\right\rvert \approx
\frac{\cos\theta_{23}}{\tan\theta_{13}}\left(1-
\frac{\dms}{\Dma}\frac{\sin2\theta_{12}}{2\tan\theta_{23}\sin\theta_{13}}\right) \approx 4.45~.\ee
As can be seen, the recent measurement of $\theta_{13}$ fixes the above ratio and it turns out to
be small due to the relatively large value of $\theta_{13}$. The ratio would have been ${\cal
O}(10)$ times larger for vanishing $\theta_{13}$. Taking the advantage of recent
measurement of $\theta_{13}$, one gets
\be \label{ratio-fixed}
\frac{{\rm BR}(\tau \rightarrow \mu \gamma)}{{\rm BR}(\mu \rightarrow e \gamma)} \approx
3.5~\frac{\bar{m}_{\se}^2}{\bar{m}_{\stau}^2}. \ee
Similarly, the other LFV decay modes are related as
\be \label{ratio-others}
\frac{{\rm BR}(\tau \rightarrow e \gamma)}{{\rm BR}(\mu \rightarrow e \gamma)} \approx
0.18~\frac{\bar{m}_{\smu}^2}{\bar{m}_{\stau}^2}~~~{\rm and}~~~
\frac{{\rm BR}(\tau \rightarrow e \gamma)}{{\rm BR}(\tau \rightarrow \mu \gamma)} \approx
0.05~\frac{\bar{m}_{\smu}^2}{\bar{m}_{\se}^2}.\ee

Note that predictions in Eqs. (\ref{ratio-fixed}, \ref{ratio-others}) do not depend on the scale of
triplet mass or on the couplings like $\lambda_{u,d}$. However they depend on the masses of the
sleptons of different generations \cite{Joaquim:2009vp}. A hierarchical slepton spectrum introduces
a ``fudge factor'' like $\bar{m}_{\se}^2/\bar{m}_{\stau}^2$ which makes the above prediction
vulnerable to the details of soft SUSY breaking parameters. In generic MSSM scenarios like
phenomenological MSSM, Eqs. (\ref{ratio-fixed}, \ref{ratio-others}) do not give any robust
prediction since the masses of different generations of sleptons are independent from each other.
However in the class of models in which some universality is assumed between the soft masses of
different generations at the high scale, $\bar{m}_{\se}$, $\bar{m}_{\smu}$ and $\bar{m}_{\stau}$ are
not completely independent and one can still obtain some useful predictions from Eqs.
(\ref{ratio-fixed}, \ref{ratio-others}). We discuss this in detail in the next section.

\section{Numerical analysis: mSUGRA}
\label{msugra}
We now discuss the charged LFV constraints on SUSY type II seesaw scenario with mSUGRA-like
boundary conditions through detailed numerical analysis. As it is well known, mSUGRA provides
the most economical set of the GUT scale soft SUSY breaking parameters that includes a
universal scalar mass $m_0$, a common gaugino mass $m_{1/2}$, a universal
trilinear coupling $A_0$ in addition to the SUSY preserving parameters $\tan\beta$ and the sign of
$\mu$ parameter. Keeping the future reach of direct search experiments like the LHC and indirect
searches from the flavor physics experiments in mind, we scan the above parameters in the following
ranges:
\begin{align} \label{prm}
 m_0 &\in [0,~5]~{\rm TeV}; \nonumber \\
 m_{1/2} &\in [0,~2]~{\rm TeV}; \nonumber \\
 A_0 &\in [-15,~15]~{\rm TeV}; \nonumber \\
 \tan\beta &\in [5,~60]; \nonumber \\
 {\rm sign}(\mu) &\in \{-,+\}. 
\end{align}
 
As discussed in the last section, we have ${\bf 15} + \overline{\bf 15}$ pair of Higgs in the 
SUSY $SU(5)$ model and we impose the following mSUGRA-like boundary condition for its soft mass
\be \label{soft_15}
m_{\overline{\bf 15}} = m_{\bf 15} = m_0 \ee
at the GUT scale. For neutrinos, we assume normal hierarchy and set
\be \label{nu-spectrum}
m_{\nu_1} = 0.001~{\rm eV};~m_{\nu_2}=\sqrt{\dms + m_{\nu_1}^2}~{\rm and}~m_{\nu_3}=\sqrt{\Dma +
m_{\nu_1}^2}\ . \ee
The RG running effects in neutrino masses and mixing angles are known to be negligible for such
hierarchical neutrinos \cite{Antusch:2003kp} and hence ${\bf Y}_T$ in Eq. (\ref{numass}) can be 
determined from the low energy data itself. For $\dms$, $\Dma$, $\theta_{12}$, $\theta_{23}$ and
$\theta_{13}$,  we use the central values obtained from the recent global fit of neutrino data
\cite{GonzalezGarcia:2012sz}. Further, we assume $\lambda_{u,d}=0.5$ throughout our analysis.

We carry out the numerical analysis using publicly available package {\tt SPheno}
\cite{Porod:2011nf}. It evaluates 2-loop RGEs for SUSY type II seesaw and incorporates full 1-loop
SUSY threshold corrections to the sparticle masses \cite{Pierce:1996zz}. Further, it calculates
complete 1-loop and dominant 2-loop corrections in the $\mu$ parameter and checks for consistency of
the radiative electroweak symmetry breaking conditions (REWSB) \cite{Dedes:2002dy} at the scale
$M_{\rm SUSY}$. Similarly, for the Higgs sector, it computes complete 1-loop and dominant 2-loop
contributions which are of $\mathcal{O}\left[ \alpha_s \left( \alpha_t +  \alpha_t \right) + \left(
\alpha_t +  \alpha_t \right)^2 + \alpha_{\tau} \alpha_b + \alpha_{\tau}^2
\right]$ \cite{Brignole:2001jy,Dedes:2002dy,Degrassi:2001yf,slavich}. While scanning, we collect
only those solutions which have (a) successful REWSB, (b) non tachyonic spectrum and (c) charge and
color neutral particle as the lightest sparticles. Next, we impose the current direct search limits
at 95\% C.L. on the masses of all sparticles given by PDG \cite{Beringer:1900zz}. We also
impose the following constraints (at 95\% C.L.) on the collected data points.
\begin{align}
\label{indirect-limits}
m_{h^0} &\in [122.5,\ 129.5]\ {\rm GeV}, \nonumber \\
{\rm BR}(B \rightarrow X_s \gamma) &\in [2.99,\ 3.87]\times 10^{-4}, \nonumber \\
{\rm BR}(B \rightarrow \tau \nu_\tau) &\in [0.44,\ 1.48]\times 10^{-4}, \nonumber \\
{\rm BR}(B_s \rightarrow \mu^+ \mu^-) &\in [0.8,\ 6.2]\times 10^{-9}, \nonumber \\
{\rm BR}(B_d \rightarrow \mu^+ \mu^-) &< 9.4 \times 10^{-10} 
\end{align}
The above range in Higgs mass includes 95\% C.L. errors from the ATLAS \cite{atlashiggs} and
CMS \cite{cmshiggs} data as well as theoretical uncertainty of about 1.5 GeV \cite{Arbey:2012dq}.
For ${\rm BR}(B \rightarrow X_s \gamma)$, we use the updated global average reported in
\cite{Amhis:2012bh}. The BELLE collaboration has recently updated their measurement of ${\rm BR}(B
\rightarrow \tau \nu_\tau)$ using hadronic tagging \cite{Adachi:2012mm} and we take its updated
global average \cite{Btaunu}. Further, we use the first measurement of ${\rm BR}(B_s \rightarrow
\mu^+ \mu^-)$ and updated limit on ${\rm BR}(B_d \rightarrow \mu^+ \mu^-)$ reported at the LHCb
\cite{Aaij:2012nna}. We do not insist here for the SUSY solution to muon $(g-2)$ discrepancy.

\begin{figure}[!ht]
\centering
\subfigure{\includegraphics[width=0.45\textwidth]{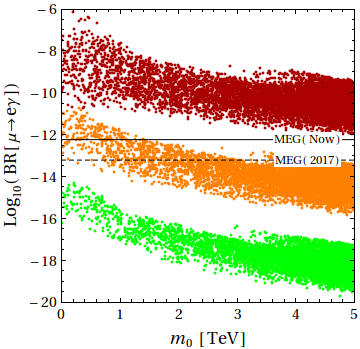}} \quad
\subfigure{\includegraphics[width=0.455\textwidth]{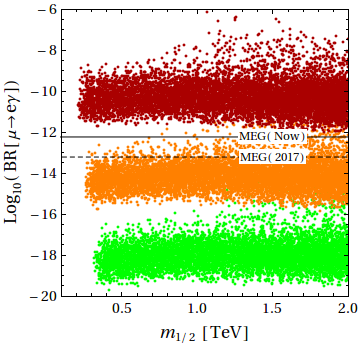}}
\caption{Constraints on $m_0$ and $m_{1/2}$ from LFV decay $\mu\rightarrow e\gamma$.
In both the panels, the red (upper), orange (middle) and green (lower) points correspond to the
triplet mass scale $M_{T}=10^{14},~10^{13}$ and $10^{12}$ GeV respectively and $\lambda_{u,d}=0.5$.
The other parameters are varied as mentioned in Eq. (\ref{prm}) and various direct and indirect
constraints are applied on the parameters as discussed in the text. The different horizontal lines
present the current limits and future sensitivities of ongoing experiments.}
\label{fig-1}
\end{figure}

We now discuss the results of numerical analysis. As already mentioned, the strongest bound on
LFVs comes from non-observation of $\mu \to e\gamma$ at the MEG experiment. In Fig. \ref{fig-1},
we show the constraints on $m_0$ and $m_{1/2}$ arising
from such a bound.
As can be seen from Fig. \ref{fig-1}, the current limit on ${\rm BR}(\mu \to e \gamma)$ rules out
completely the low values of soft masses, {\it i.e.} $m_0 < 5$ TeV and $m_{1/2} < 2$ TeV, for 
$M_T \gtrsim 10^{14}$ GeV making them inaccessible at the LHC. 
We also note that narrowing down the Higgs mass range to 124-128 GeV eliminates some
points without clearly disfavoring any particular region in Fig. \ref{fig-1}.
The BR decreases for smaller
$M_T$ as discussed in Section \ref{sect-2a}. The near future limit expected from the updated MEG
\cite{Adam:2013mnn} can further constrain the low $m_0$-$m_{1/2}$ region for $M_T>10^{13}$ GeV
however it does not put any constraint on the soft SUSY breaking parameters if the triplet mass
scale is below $10^{13}$ GeV.

\begin{figure}[!ht] 
 \centering
\subfigure{\includegraphics[width=0.45\textwidth]{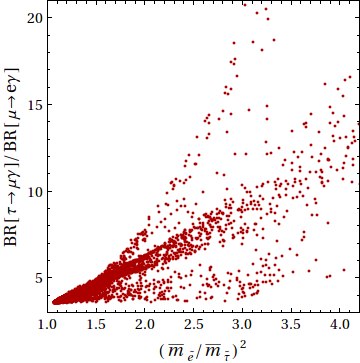}} \quad
\subfigure{\includegraphics[width=0.45\textwidth]{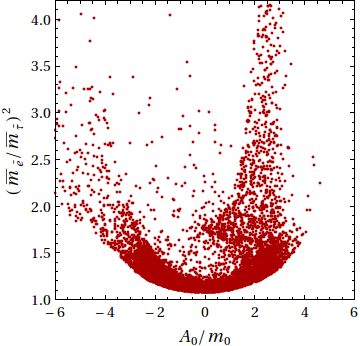}}
\caption{The left panel displays the ratio ${\rm BR}(\tau \rightarrow \mu \gamma)/{\rm BR}(\mu
\rightarrow e \gamma)$  as a function of fudge factor $\bar{m}_{\se}^2/\bar{m}_{\stau}^2$. The
correlation between the fudge factor and $A_0/m_0$ is shown in the right panel.}
\label{fig-2}
\end{figure}

\begin{figure}[!ht]
\centering
\subfigure{\includegraphics[width=0.45\textwidth]{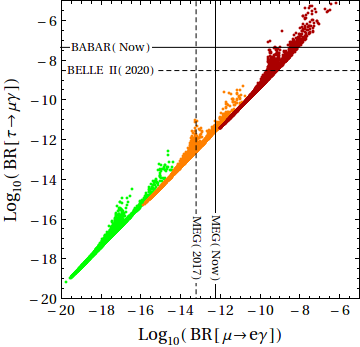}} \quad
\subfigure{\includegraphics[width=0.45\textwidth]{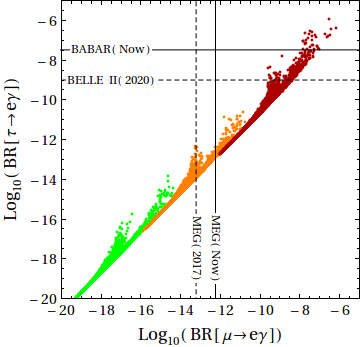}}
\caption{Correlations between different LFV decays. In both the panels, the red (upper), orange
(middle) and green (lower) points correspond to the triplet mass scale $M_{T}=10^{14},\ 10^{13}$ and
$10^{12}$ GeV respectively and $\lambda_{u,d}=0.5$. The other parameters are varied as mentioned in
Eq. (\ref{prm}) and various direct and indirect constraints are applied on the parameters as
discussed in the text. The different horizontal and vertical lines present the current limits and
future sensitivities of ongoing experiments.}
\label{fig-3}
\end{figure}

As discussed earlier in Section \ref{sect-2b}, the ratio of branching ratios of $\tau \to \mu
\gamma$ and $\mu \to e \gamma$ is fixed up to a fudge factor after the precisely known value of
$\theta_{13}$. The correlation between these two LFV channels as a function of the fudge
factor is displayed in Fig. \ref{fig-2}.
The large trilinear coupling $|A_0| \gg m_0$ together with large $\tan\beta$ can induce
significant splittings between the masses of third and first two generations of sfermions even in
the class of models with universal boundary conditions. The off-diagonal term in the effective
$2\times 2$ mass matrix of an $i^{\rm th}$ generation sfermions is proportional to $A_0 y_i$. Thus
its contribution to the masses of third generation sfermions is significant compared to the first
two generations. In the case of sleptons, this makes staus lighter compared to that of first two
generations of sleptons. This is reflected in Fig. \ref{fig-2} where large values of $A_0$ drives
the ratio $\bar{m}_{\se}/\bar{m}_{\stau}$ greater than unity. We also note that the hierarchy
between $\bar{m}_{\se}$ and $\bar{m}_{\stau}$ becomes stronger for large values of $\tan\beta$. As a
result, one gets relatively enhanced values of ${\rm BR}(\tau \rightarrow \mu \gamma)/{\rm
BR}(\mu \rightarrow e \gamma)$. Further, note that $|A_0|/m_0$ cannot be arbitrarily large since
such values correspond to tachyonic spectrum for the third generation squarks and sleptons in
mSUGRA-like models. It follows from Fig. \ref{fig-2} that
\be \label{tmg-bound}
3.5 \lesssim \frac{{\rm BR}(\tau \rightarrow \mu \gamma)}{{\rm BR}(\mu \rightarrow e \gamma)}
\lesssim 20~. \ee
The above prediction is the distinctive feature of the type II seesaw mechanism in the models with
mSUGRA-like boundary conditions and it does not depend on the other parameters of the model. This is
more clearly shown in Fig. \ref{fig-3}, in which we also show the correlation among the branching
ratios of $\mu \to e \gamma$ and $\tau \to e \gamma$. 
As can be seen from Fig. \ref{fig-3}, the current upper limit on ${\rm BR}(\mu \rightarrow e
\gamma)$ implies an upper limit ${\rm BR}(\tau \rightarrow \mu \gamma) \lesssim 10^{-11}$
which is significantly smaller than the sensitivity of current generation experiments. Thus, any
signal of $\tau \rightarrow \mu \gamma$ (or $\tau \rightarrow e \gamma$) in near future would rule
out the type II seesaw scenario discussed here. We also compute the other charged LFV decays like 
${\rm BR}(l_i \rightarrow 3 l_j)$. However we do not show their plots here because they are found in
good agreement with the approximate relation \cite{Arganda:2005ji}
\be \label{clfv-3body}
\frac{{\rm BR}(l_j \to 3 l_i)}{{\rm BR}(l_j \to l_i \gamma)}\approx\frac{2\alpha}{3 \pi} \left(
{\rm log} \left( \frac{m_{l_j}}{m_{l_i}}\right) - \frac{11}{8}\right)~. \ee

\begin{figure}[!ht]
\centering
\includegraphics[width=0.5\textwidth]{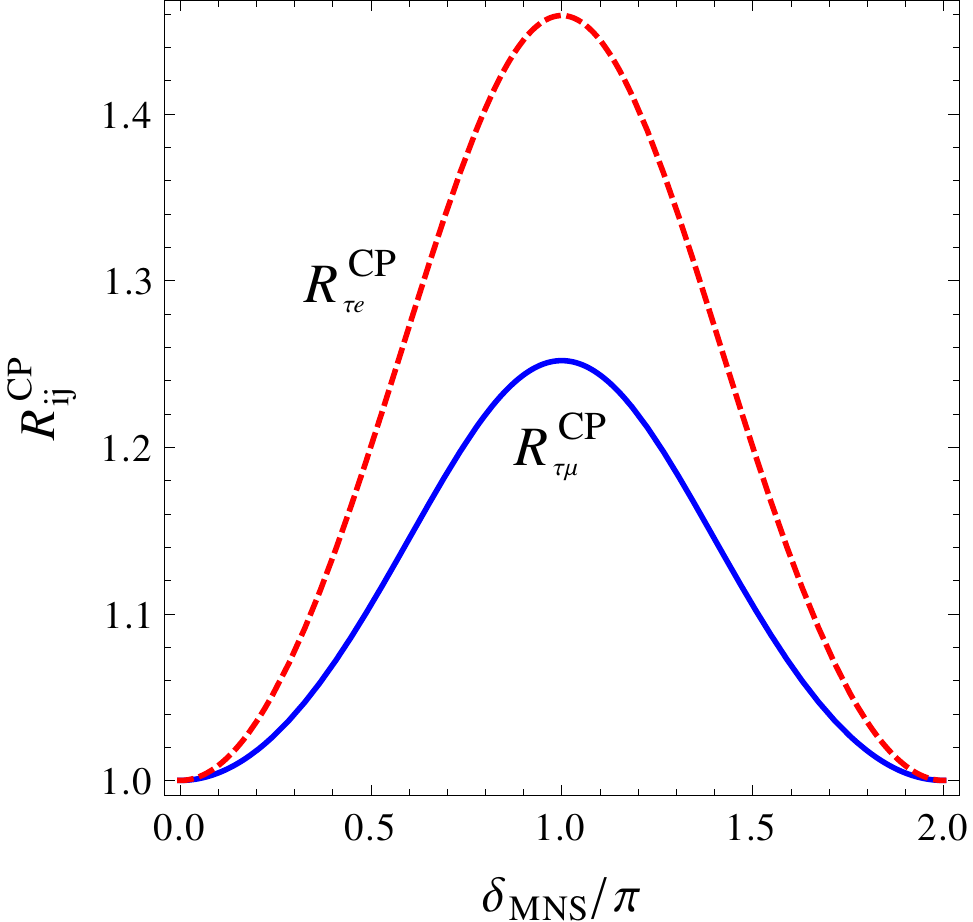}
\caption{The ratio $R_{ij}^{\rm CP} \equiv \dfrac{\left\lvert(\mathbf{m}^2_{\tilde
L})_{ij}/(\mathbf{m}^2_{\tilde L})_{\mu e}\right\rvert_{\delta_{MNS} \neq 0}
}{\left\lvert(\mathbf{m}^2_{\tilde
L})_{ij}/(\mathbf{m}^2_{\tilde L})_{\mu e}\right\rvert_{\delta_{MNS} = 0}}$
as a function of Dirac CP phase. The solid (blue) and dashed (red) lines correspond to $ij = \tau
\mu$ and $ij = \tau e$ respectively. }
\label{fig-4}
\end{figure}

Before we end this section, we comment on the possible effects of Dirac CP violation on the above
results. Note that we neglected the Dirac CP phase $\delta_{\rm MNS}$ in the lepton sector
while deriving the ratio in Eq. (\ref{ratio-ml}). The nonzero $\delta_{\rm MNS}$ can modify it as
displayed in Fig. \ref{fig-4}.
Clearly, the ratio can get enhanced by 25\% if $\delta_{\rm MNS}=\pi$. As a result, the prediction
for ${\rm BR}(\tau \rightarrow \mu \gamma)/{\rm BR}(\mu \rightarrow e \gamma)$ in Eq.
(\ref{tmg-bound}) can increase at most by 50\% for nonzero CP violation in the lepton sector.
We perform a numerical analysis taking arbitrary CP violation into account and find
a more conservative range
\be \label{tmg-bound+CP}
3.5 \lesssim \frac{{\rm BR}(\tau \rightarrow \mu \gamma)}{{\rm BR}(\mu \rightarrow e \gamma)}
\lesssim 30~. \ee
The effect of CP violation is larger on the ratio ${\rm BR}(\tau \rightarrow e \gamma)/{\rm BR}(\mu
\rightarrow e \gamma)$ as can be seen in Fig. \ref{fig-4}.

\section{Numerical analysis: non-universal triplet scalar mass}
\label{nutm}
We now discuss the cancellations in the LFV that arise form the non-universal masses of
triplet scalar. In this case, one can rewrite the flavor violations in the slepton sector shown in
Eq. (\ref{Delta1}) as 
\be \label{Delta-nuthm}
\lrb \mathbf{m}^2_{\tilde L} \rrb_{ij} \approx -\frac{3(2m_0^2+ m_T^2 +A_0^2)}{8 \pi^2} ( \mathbf{Y}_T^\dagger \mathbf{Y}_T )_{ij}
\log\left( \frac{M_{\rm GUT}}{M_T}\right), \ee
where $m_T\equiv m_{\bf 15}$ is the GUT scale value of soft mass of triplet field residing in the
{\bf 15}-plet scalar of $SU(5)$. Since the MSSM doublets reside in the ${\bf 5}$ and $\overline{\bf 5}$
representation of $SU(5)$, its soft mass $m_0$ can be different form $m_{\bf 15}$ at the GUT scale as
dictated by the gauge invariance. Even in the case of universal scalar mass at the Planck scale,
the splitting between $m_0$ and $m_{\bf 15}$ gets induced at the GUT scale due to different RG 
running of ${\bf 15}$-plet and ${\bf 5}$-plet from the Planck scale to the GUT scale. In such cases,
one naturally expects $m_{\bf 15}^2 \neq m_0^2$ and depending on their relative magnitude of $m_{\bf 15}$ the
flavor violation gets enhanced or reduced in comparison to mSUGRA scenario, as it can
be seen from Eq. (\ref{Delta-nuthm}). The choice $m_{\bf 15}^2 < m_0^2$ decreases the magnitude of
lepton flavor violation so it can relax the MEG constraint on the model.

\begin{figure}[ht!]
\centering
\subfigure{\includegraphics[width=0.45\textwidth]{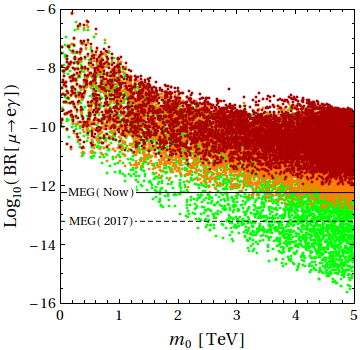}} \quad
\subfigure{\includegraphics[width=0.45\textwidth]{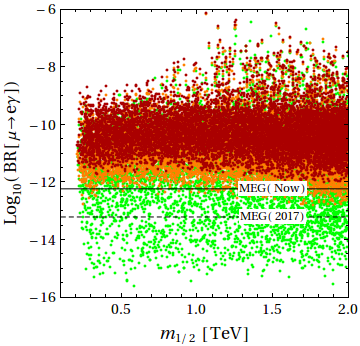}}
\caption{Constraints on $m_0$ and $m_{1/2}$ from LFV decay $\mu\rightarrow e\gamma$ in
NUTM. In both the panels, the red (upper), orange (middle) and green (lower) points
correspond to the triplet soft mass $m_{\bf 15}^2 = m_0^2, -m_0^2$ and $-2m_0^2$ respectively and the same
values of $M_T=10^{14}$ GeV and $\lambda_{u,d}=0.5$. The other parameters are varied as mentioned in
Eq.
(\ref{prm}) and various direct and indirect constraints are applied on the parameters as
discussed in the text. The different horizontal lines present the current limits and future
sensitivities of ongoing experiments.}
\label{fig-5}
\end{figure}

We demonstrate this using the same kind of numerical analysis performed in the last section but with
non-universal soft mass for the triplet scalar. For example, we study three different cases
corresponding to $m_{\bf 15}^2=\lcb m_0^2,\ -m_0^2,\ -2m_0^2 \rcb$ for the same triplet scale
$M_T=10^{14}$ GeV. The results are displayed in Fig. \ref{fig-5}.
As can be seen, the negative $m_{\bf 15}^2$ induces the cancellations between soft masses and
significantly reduces BR($\mu \to e \gamma$). One can see that the current MEG bound on $\mu \to e
\gamma$ still provides powerful constraints on $m_0$ and $m_{1/2}$ as long as $m_{\bf 15}^2>-m_0^2$.
In case of $m_{\bf 15}^2=-2 m_0^2$, the MEG constraint does not put any restrictions on $m_{1/2}$
and also allows $m_0$ as low as 1.5 TeV.

\begin{figure}[ht!]
\centering
\subfigure{\includegraphics[width=0.45\textwidth]{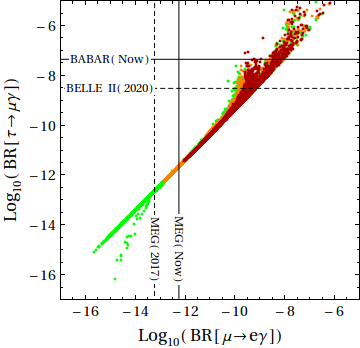}} \quad
\subfigure{\includegraphics[width=0.45\textwidth]{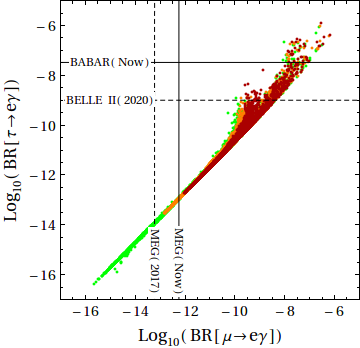}}
\caption{Correlations between different LFV decays in NUTM. In both the panels, the red
(upper), orange (middle) and green (lower) points correspond to the triplet soft mass
$m_{\bf 15}^2 = m_0^2, -m_0^2$ and $-2m_0^2$ respectively and the same value of $M_T=10^{14}$ GeV and
$\lambda_{u,d}=0.5$. The other parameters are varied as mentioned in Eq.
(\ref{prm}) and various direct and indirect constraints are applied on the parameters as
discussed in the text. The different horizontal and vertical lines present the current limits and
future sensitivities of ongoing experiments.}
\label{fig-6}
\end{figure}

We also show the correlations between the branching ratios of different LFV decays in Fig.
\ref{fig-6}. 
It is important to note that the ratio $\lrb \mathbf{m}^2_{\tilde L} \rrb_{\tau \mu}/ \lrb \mathbf{m}^2_{\tilde L} 
\rrb_{\mu e}$ does not get modified in NUTM as can be seen from
Eq. (\ref{Delta-nuthm}). Also, the non-universality $m_{\bf 15}^2 \neq m_0^2$ has very tiny effects
on the fudge factor $\bar{m}_{\se}^2/\bar{m}_{\stau}^2$ as they are induced only through the RG
running. As a result, Eq. (\ref{tmg-bound}) obtained in the mSUGRA case also holds true in this case
as can be seen from Fig. \ref{fig-6}.


\section{Summary}
\label{summary}

We revisit supersymmetric type II seesaw mechanism and present an updated analysis of the charged
lepton flavor violations arise in this model. We show that in CMSSM/mSUGRA like models, the
present experimental limit on BR($\mu \to e \gamma$) disfavors the soft SUSY breaking parameters
$m_0<5$ TeV and $m_{1/2}<2$ TeV if the triplet Yukawas are of ${\cal O}$(1). This corresponds to a
SUSY particle spectrum which is beyond the reach of the LHC. LFV constraint on SUSY spectrum becomes
milder if the Yukawas are small or, in other words, the mass of triplet scalar is below $10^{13}$
GeV. We show that interesting cancellation in the magnitude of charged LFVs arise if the
universality condition is relaxed for the soft mass of triplet scalar. In such a
case, MEG constraint can be evaded up to certain extent which allows relatively light SUSY
spectrum.

We show that the recent observation of $\theta_{13}$ fixes ratios of decay rates of various
charged LFV channels in a class of SUSY type II seesaw models in which the slepton
masses are universal at the GUT scale. These ratios depend on the leptonic Dirac CP phase and
on the details of soft SUSY breaking parameters and $\tan\beta$. Taking all the uncertainty factors
into account, the mSUGRA/CMSSM and NUTM models discussed here predict ${\rm BR}(\tau\rightarrow \mu
\gamma)/{\rm BR}(\mu \rightarrow e \gamma) \in [3.5,~30]$. This prediction distinguishes
type II seesaw from the other variants of seesaw mechanism. Any observational evidence of the
deviation from this prediction can rule out type II seesaw mechanism in these models as an
only mechanism to explain the smallness of neutrino masses.

\begin{acknowledgments}
We are grateful to Sudhir K. Vempati for encouragement and fruitful discussions throughout the
course of this work. We also thank him for reading the manuscript carefully. D. C. would like to
thank Amol Dighe for useful discussions and for the hospitality and support for his visit to Tata
Institute of Fundamental Research where part of this work was carried out. K. M. P. acknowledges the
hospitality of Centre for High Energy Physics, Indian Institute of Science where this work was
initiated.
\end{acknowledgments}

\bibliography{refs-typeII}

\begin{thebibliography}{66}%
\makeatletter
\providecommand \@ifxundefined [1]{%
 \@ifx{#1\undefined}
}%
\providecommand \@ifnum [1]{%
 \ifnum #1\expandafter \@firstoftwo
 \else \expandafter \@secondoftwo
 \fi
}%
\providecommand \@ifx [1]{%
 \ifx #1\expandafter \@firstoftwo
 \else \expandafter \@secondoftwo
 \fi
}%
\providecommand \natexlab [1]{#1}%
\providecommand \enquote  [1]{``#1''}%
\providecommand \bibnamefont  [1]{#1}%
\providecommand \bibfnamefont [1]{#1}%
\providecommand \citenamefont [1]{#1}%
\providecommand \href@noop [0]{\@secondoftwo}%
\providecommand \href [0]{\begingroup \@sanitize@url \@href}%
\providecommand \@href[1]{\@@startlink{#1}\@@href}%
\providecommand \@@href[1]{\endgroup#1\@@endlink}%
\providecommand \@sanitize@url [0]{\catcode `\\12\catcode `\$12\catcode
  `\&12\catcode `\#12\catcode `\^12\catcode `\_12\catcode `\%12\relax}%
\providecommand \@@startlink[1]{}%
\providecommand \@@endlink[0]{}%
\providecommand \url  [0]{\begingroup\@sanitize@url \@url }%
\providecommand \@url [1]{\endgroup\@href {#1}{\urlprefix }}%
\providecommand \urlprefix  [0]{URL }%
\providecommand \Eprint [0]{\href }%
\providecommand \doibase [0]{http://dx.doi.org/}%
\providecommand \selectlanguage [0]{\@gobble}%
\providecommand \bibinfo  [0]{\@secondoftwo}%
\providecommand \bibfield  [0]{\@secondoftwo}%
\providecommand \translation [1]{[#1]}%
\providecommand \BibitemOpen [0]{}%
\providecommand \bibitemStop [0]{}%
\providecommand \bibitemNoStop [0]{.\EOS\space}%
\providecommand \EOS [0]{\spacefactor3000\relax}%
\providecommand \BibitemShut  [1]{\csname bibitem#1\endcsname}%
\let\auto@bib@innerbib\@empty
\bibitem [{\citenamefont {Minkowski}(1977)}]{Minkowski:1977sc}%
  \BibitemOpen
  \bibfield  {author} {\bibinfo {author} {\bibfnamefont {P.}~\bibnamefont
  {Minkowski}},\ }\href {\doibase 10.1016/0370-2693(77)90435-X} {\bibfield
  {journal} {\bibinfo  {journal} {Phys. Lett.}\ }\textbf {\bibinfo {volume}
  {B67}},\ \bibinfo {pages} {421} (\bibinfo {year} {1977})}\BibitemShut
  {NoStop}%
\bibitem [{\citenamefont {Yanagida}()}]{Yanagida:1979as}%
  \BibitemOpen
  \bibfield  {author} {\bibinfo {author} {\bibfnamefont {T.}~\bibnamefont
  {Yanagida}},\ }\href@noop {} {\ }\bibinfo {note} {In Proceedings of the
  Workshop on the Baryon Number of the Universe and Unified Theories, Tsukuba,
  Japan, 13-14 Feb 1979}\BibitemShut {NoStop}%
\bibitem [{\citenamefont {Gell-Mann}\ \emph {et~al.}()\citenamefont
  {Gell-Mann}, \citenamefont {Ramond},\ and\ \citenamefont
  {Slansky}}]{GellMann:1980vs}%
  \BibitemOpen
  \bibfield  {author} {\bibinfo {author} {\bibfnamefont {M.}~\bibnamefont
  {Gell-Mann}}, \bibinfo {author} {\bibfnamefont {P.}~\bibnamefont {Ramond}}, \
  and\ \bibinfo {author} {\bibfnamefont {R.}~\bibnamefont {Slansky}},\
  }\href@noop {} {\ }\bibinfo {note} {Prepared for Supergravity Workshop, Stony
  Brook, New York, 27-28 Sep 1979}\BibitemShut {NoStop}%
\bibitem [{\citenamefont {Mohapatra}\ and\ \citenamefont
  {Senjanovic}(1980)}]{Mohapatra:1979ia}%
  \BibitemOpen
  \bibfield  {author} {\bibinfo {author} {\bibfnamefont {R.~N.}\ \bibnamefont
  {Mohapatra}}\ and\ \bibinfo {author} {\bibfnamefont {G.}~\bibnamefont
  {Senjanovic}},\ }\href {\doibase 10.1103/PhysRevLett.44.912} {\bibfield
  {journal} {\bibinfo  {journal} {Phys. Rev. Lett.}\ }\textbf {\bibinfo
  {volume} {44}},\ \bibinfo {pages} {912} (\bibinfo {year} {1980})}\BibitemShut
  {NoStop}%
\bibitem [{\citenamefont {Schechter}\ and\ \citenamefont
  {Valle}(1980)}]{Schechter:1980gr}%
  \BibitemOpen
  \bibfield  {author} {\bibinfo {author} {\bibfnamefont {J.}~\bibnamefont
  {Schechter}}\ and\ \bibinfo {author} {\bibfnamefont {J.}~\bibnamefont
  {Valle}},\ }\href {\doibase 10.1103/PhysRevD.22.2227} {\bibfield  {journal}
  {\bibinfo  {journal} {Phys.Rev.}\ }\textbf {\bibinfo {volume} {D22}},\
  \bibinfo {pages} {2227} (\bibinfo {year} {1980})}\BibitemShut {NoStop}%
\bibitem [{\citenamefont {Mohapatra}\ and\ \citenamefont
  {Senjanovic}(1981)}]{Mohapatra:1980yp}%
  \BibitemOpen
  \bibfield  {author} {\bibinfo {author} {\bibfnamefont {R.~N.}\ \bibnamefont
  {Mohapatra}}\ and\ \bibinfo {author} {\bibfnamefont {G.}~\bibnamefont
  {Senjanovic}},\ }\href {\doibase 10.1103/PhysRevD.23.165} {\bibfield
  {journal} {\bibinfo  {journal} {Phys.Rev.}\ }\textbf {\bibinfo {volume}
  {D23}},\ \bibinfo {pages} {165} (\bibinfo {year} {1981})}\BibitemShut
  {NoStop}%
\bibitem [{\citenamefont {Lazarides}\ \emph {et~al.}(1981)\citenamefont
  {Lazarides}, \citenamefont {Shafi},\ and\ \citenamefont
  {Wetterich}}]{Lazarides:1980nt}%
  \BibitemOpen
  \bibfield  {author} {\bibinfo {author} {\bibfnamefont {G.}~\bibnamefont
  {Lazarides}}, \bibinfo {author} {\bibfnamefont {Q.}~\bibnamefont {Shafi}}, \
  and\ \bibinfo {author} {\bibfnamefont {C.}~\bibnamefont {Wetterich}},\ }\href
  {\doibase 10.1016/0550-3213(81)90354-0} {\bibfield  {journal} {\bibinfo
  {journal} {Nucl.Phys.}\ }\textbf {\bibinfo {volume} {B181}},\ \bibinfo
  {pages} {287} (\bibinfo {year} {1981})}\BibitemShut {NoStop}%
\bibitem [{\citenamefont {Cheng}\ and\ \citenamefont
  {Li}(1980)}]{Cheng:1980qt}%
  \BibitemOpen
  \bibfield  {author} {\bibinfo {author} {\bibfnamefont {T.}~\bibnamefont
  {Cheng}}\ and\ \bibinfo {author} {\bibfnamefont {L.-F.}\ \bibnamefont {Li}},\
  }\href {\doibase 10.1103/PhysRevD.22.2860} {\bibfield  {journal} {\bibinfo
  {journal} {Phys.Rev.}\ }\textbf {\bibinfo {volume} {D22}},\ \bibinfo {pages}
  {2860} (\bibinfo {year} {1980})}\BibitemShut {NoStop}%
\bibitem [{\citenamefont {Foot}\ \emph {et~al.}(1989)\citenamefont {Foot},
  \citenamefont {Lew}, \citenamefont {He},\ and\ \citenamefont
  {Joshi}}]{Foot:1988aq}%
  \BibitemOpen
  \bibfield  {author} {\bibinfo {author} {\bibfnamefont {R.}~\bibnamefont
  {Foot}}, \bibinfo {author} {\bibfnamefont {H.}~\bibnamefont {Lew}}, \bibinfo
  {author} {\bibfnamefont {X.}~\bibnamefont {He}}, \ and\ \bibinfo {author}
  {\bibfnamefont {G.~C.}\ \bibnamefont {Joshi}},\ }\href {\doibase
  10.1007/BF01415558} {\bibfield  {journal} {\bibinfo  {journal} {Z.Phys.}\
  }\textbf {\bibinfo {volume} {C44}},\ \bibinfo {pages} {441} (\bibinfo {year}
  {1989})}\BibitemShut {NoStop}%
\bibitem [{\citenamefont {Ma}(1998)}]{Ma:1998dn}%
  \BibitemOpen
  \bibfield  {author} {\bibinfo {author} {\bibfnamefont {E.}~\bibnamefont
  {Ma}},\ }\href {\doibase 10.1103/PhysRevLett.81.1171} {\bibfield  {journal}
  {\bibinfo  {journal} {Phys.Rev.Lett.}\ }\textbf {\bibinfo {volume} {81}},\
  \bibinfo {pages} {1171} (\bibinfo {year} {1998})},\ \Eprint
  {http://arxiv.org/abs/hep-ph/9805219} {arXiv:hep-ph/9805219 [hep-ph]}
  \BibitemShut {NoStop}%
\bibitem [{\citenamefont {Lee}(1984{\natexlab{a}})}]{Lee:1984kr}%
  \BibitemOpen
  \bibfield  {author} {\bibinfo {author} {\bibfnamefont {I.-H.}\ \bibnamefont
  {Lee}},\ }\href {\doibase 10.1016/0370-2693(84)91885-9} {\bibfield  {journal}
  {\bibinfo  {journal} {Phys.Lett.}\ }\textbf {\bibinfo {volume} {B138}},\
  \bibinfo {pages} {121} (\bibinfo {year} {1984}{\natexlab{a}})}\BibitemShut
  {NoStop}%
\bibitem [{\citenamefont {Lee}(1984{\natexlab{b}})}]{Lee:1984tn}%
  \BibitemOpen
  \bibfield  {author} {\bibinfo {author} {\bibfnamefont {I.-H.}\ \bibnamefont
  {Lee}},\ }\href {\doibase 10.1016/0550-3213(84)90117-2} {\bibfield  {journal}
  {\bibinfo  {journal} {Nucl.Phys.}\ }\textbf {\bibinfo {volume} {B246}},\
  \bibinfo {pages} {120} (\bibinfo {year} {1984}{\natexlab{b}})}\BibitemShut
  {NoStop}%
\bibitem [{\citenamefont {Borzumati}\ and\ \citenamefont
  {Masiero}(1986)}]{Borzumati:1986qx}%
  \BibitemOpen
  \bibfield  {author} {\bibinfo {author} {\bibfnamefont {F.}~\bibnamefont
  {Borzumati}}\ and\ \bibinfo {author} {\bibfnamefont {A.}~\bibnamefont
  {Masiero}},\ }\href {\doibase 10.1103/PhysRevLett.57.961} {\bibfield
  {journal} {\bibinfo  {journal} {Phys.Rev.Lett.}\ }\textbf {\bibinfo {volume}
  {57}},\ \bibinfo {pages} {961} (\bibinfo {year} {1986})}\BibitemShut
  {NoStop}%
\bibitem [{\citenamefont {Hall}\ \emph {et~al.}(1986)\citenamefont {Hall},
  \citenamefont {Kostelecky},\ and\ \citenamefont {Raby}}]{Hall:1985dx}%
  \BibitemOpen
  \bibfield  {author} {\bibinfo {author} {\bibfnamefont {L.~J.}\ \bibnamefont
  {Hall}}, \bibinfo {author} {\bibfnamefont {V.~A.}\ \bibnamefont
  {Kostelecky}}, \ and\ \bibinfo {author} {\bibfnamefont {S.}~\bibnamefont
  {Raby}},\ }\href {\doibase 10.1016/0550-3213(86)90397-4} {\bibfield
  {journal} {\bibinfo  {journal} {Nucl.Phys.}\ }\textbf {\bibinfo {volume}
  {B267}},\ \bibinfo {pages} {415} (\bibinfo {year} {1986})}\BibitemShut
  {NoStop}%
\bibitem [{\citenamefont {Gabbiani}\ and\ \citenamefont
  {Masiero}(1989)}]{Gabbiani:1988rb}%
  \BibitemOpen
  \bibfield  {author} {\bibinfo {author} {\bibfnamefont {F.}~\bibnamefont
  {Gabbiani}}\ and\ \bibinfo {author} {\bibfnamefont {A.}~\bibnamefont
  {Masiero}},\ }\href {\doibase 10.1016/0550-3213(89)90492-6} {\bibfield
  {journal} {\bibinfo  {journal} {Nucl.Phys.}\ }\textbf {\bibinfo {volume}
  {B322}},\ \bibinfo {pages} {235} (\bibinfo {year} {1989})}\BibitemShut
  {NoStop}%
\bibitem [{\citenamefont {Hisano}\ \emph {et~al.}(1995)\citenamefont {Hisano},
  \citenamefont {Moroi}, \citenamefont {Tobe}, \citenamefont {Yamaguchi},\ and\
  \citenamefont {Yanagida}}]{Hisano:1995nq}%
  \BibitemOpen
  \bibfield  {author} {\bibinfo {author} {\bibfnamefont {J.}~\bibnamefont
  {Hisano}}, \bibinfo {author} {\bibfnamefont {T.}~\bibnamefont {Moroi}},
  \bibinfo {author} {\bibfnamefont {K.}~\bibnamefont {Tobe}}, \bibinfo {author}
  {\bibfnamefont {M.}~\bibnamefont {Yamaguchi}}, \ and\ \bibinfo {author}
  {\bibfnamefont {T.}~\bibnamefont {Yanagida}},\ }\href {\doibase
  10.1016/0370-2693(95)00954-J} {\bibfield  {journal} {\bibinfo  {journal}
  {Phys.Lett.}\ }\textbf {\bibinfo {volume} {B357}},\ \bibinfo {pages} {579}
  (\bibinfo {year} {1995})},\ \Eprint {http://arxiv.org/abs/hep-ph/9501407}
  {arXiv:hep-ph/9501407 [hep-ph]} \BibitemShut {NoStop}%
\bibitem [{\citenamefont {Hisano}\ \emph {et~al.}(1996)\citenamefont {Hisano},
  \citenamefont {Moroi}, \citenamefont {Tobe},\ and\ \citenamefont
  {Yamaguchi}}]{Hisano:1995cp}%
  \BibitemOpen
  \bibfield  {author} {\bibinfo {author} {\bibfnamefont {J.}~\bibnamefont
  {Hisano}}, \bibinfo {author} {\bibfnamefont {T.}~\bibnamefont {Moroi}},
  \bibinfo {author} {\bibfnamefont {K.}~\bibnamefont {Tobe}}, \ and\ \bibinfo
  {author} {\bibfnamefont {M.}~\bibnamefont {Yamaguchi}},\ }\href {\doibase
  10.1103/PhysRevD.53.2442} {\bibfield  {journal} {\bibinfo  {journal}
  {Phys.Rev.}\ }\textbf {\bibinfo {volume} {D53}},\ \bibinfo {pages} {2442}
  (\bibinfo {year} {1996})},\ \Eprint {http://arxiv.org/abs/hep-ph/9510309}
  {arXiv:hep-ph/9510309 [hep-ph]} \BibitemShut {NoStop}%
\bibitem [{\citenamefont {Hisano}\ and\ \citenamefont
  {Nomura}(1999)}]{Hisano:1998fj}%
  \BibitemOpen
  \bibfield  {author} {\bibinfo {author} {\bibfnamefont {J.}~\bibnamefont
  {Hisano}}\ and\ \bibinfo {author} {\bibfnamefont {D.}~\bibnamefont
  {Nomura}},\ }\href {\doibase 10.1103/PhysRevD.59.116005} {\bibfield
  {journal} {\bibinfo  {journal} {Phys.Rev.}\ }\textbf {\bibinfo {volume}
  {D59}},\ \bibinfo {pages} {116005} (\bibinfo {year} {1999})},\ \Eprint
  {http://arxiv.org/abs/hep-ph/9810479} {arXiv:hep-ph/9810479 [hep-ph]}
  \BibitemShut {NoStop}%
\bibitem [{\citenamefont {Masiero}\ \emph {et~al.}(2003)\citenamefont
  {Masiero}, \citenamefont {Vempati},\ and\ \citenamefont
  {Vives}}]{Masiero:2002jn}%
  \BibitemOpen
  \bibfield  {author} {\bibinfo {author} {\bibfnamefont {A.}~\bibnamefont
  {Masiero}}, \bibinfo {author} {\bibfnamefont {S.~K.}\ \bibnamefont
  {Vempati}}, \ and\ \bibinfo {author} {\bibfnamefont {O.}~\bibnamefont
  {Vives}},\ }\href {\doibase 10.1016/S0550-3213(02)01031-3} {\bibfield
  {journal} {\bibinfo  {journal} {Nucl.Phys.}\ }\textbf {\bibinfo {volume}
  {B649}},\ \bibinfo {pages} {189} (\bibinfo {year} {2003})},\ \Eprint
  {http://arxiv.org/abs/hep-ph/0209303} {arXiv:hep-ph/0209303 [hep-ph]}
  \BibitemShut {NoStop}%
\bibitem [{\citenamefont {Babu}\ \emph {et~al.}(2003)\citenamefont {Babu},
  \citenamefont {Dutta},\ and\ \citenamefont {Mohapatra}}]{Babu:2002tb}%
  \BibitemOpen
  \bibfield  {author} {\bibinfo {author} {\bibfnamefont {K.}~\bibnamefont
  {Babu}}, \bibinfo {author} {\bibfnamefont {B.}~\bibnamefont {Dutta}}, \ and\
  \bibinfo {author} {\bibfnamefont {R.}~\bibnamefont {Mohapatra}},\ }\href
  {\doibase 10.1103/PhysRevD.67.076006} {\bibfield  {journal} {\bibinfo
  {journal} {Phys.Rev.}\ }\textbf {\bibinfo {volume} {D67}},\ \bibinfo {pages}
  {076006} (\bibinfo {year} {2003})},\ \Eprint
  {http://arxiv.org/abs/hep-ph/0211068} {arXiv:hep-ph/0211068 [hep-ph]}
  \BibitemShut {NoStop}%
\bibitem [{\citenamefont {Masiero}\ \emph {et~al.}(2004)\citenamefont
  {Masiero}, \citenamefont {Vempati},\ and\ \citenamefont
  {Vives}}]{Masiero:2004js}%
  \BibitemOpen
  \bibfield  {author} {\bibinfo {author} {\bibfnamefont {A.}~\bibnamefont
  {Masiero}}, \bibinfo {author} {\bibfnamefont {S.~K.}\ \bibnamefont
  {Vempati}}, \ and\ \bibinfo {author} {\bibfnamefont {O.}~\bibnamefont
  {Vives}},\ }\href {\doibase 10.1088/1367-2630/6/1/202} {\bibfield  {journal}
  {\bibinfo  {journal} {New J.Phys.}\ }\textbf {\bibinfo {volume} {6}},\
  \bibinfo {pages} {202} (\bibinfo {year} {2004})},\ \Eprint
  {http://arxiv.org/abs/hep-ph/0407325} {arXiv:hep-ph/0407325 [hep-ph]}
  \BibitemShut {NoStop}%
\bibitem [{\citenamefont {Calibbi}\ \emph {et~al.}(2006)\citenamefont
  {Calibbi}, \citenamefont {Faccia}, \citenamefont {Masiero},\ and\
  \citenamefont {Vempati}}]{Calibbi:2006nq}%
  \BibitemOpen
  \bibfield  {author} {\bibinfo {author} {\bibfnamefont {L.}~\bibnamefont
  {Calibbi}}, \bibinfo {author} {\bibfnamefont {A.}~\bibnamefont {Faccia}},
  \bibinfo {author} {\bibfnamefont {A.}~\bibnamefont {Masiero}}, \ and\
  \bibinfo {author} {\bibfnamefont {S.}~\bibnamefont {Vempati}},\ }\href
  {\doibase 10.1103/PhysRevD.74.116002} {\bibfield  {journal} {\bibinfo
  {journal} {Phys.Rev.}\ }\textbf {\bibinfo {volume} {D74}},\ \bibinfo {pages}
  {116002} (\bibinfo {year} {2006})},\ \Eprint
  {http://arxiv.org/abs/hep-ph/0605139} {arXiv:hep-ph/0605139 [hep-ph]}
  \BibitemShut {NoStop}%
\bibitem [{\citenamefont {Antusch}\ \emph {et~al.}(2006)\citenamefont
  {Antusch}, \citenamefont {Arganda}, \citenamefont {Herrero},\ and\
  \citenamefont {Teixeira}}]{Antusch:2006vw}%
  \BibitemOpen
  \bibfield  {author} {\bibinfo {author} {\bibfnamefont {S.}~\bibnamefont
  {Antusch}}, \bibinfo {author} {\bibfnamefont {E.}~\bibnamefont {Arganda}},
  \bibinfo {author} {\bibfnamefont {M.}~\bibnamefont {Herrero}}, \ and\
  \bibinfo {author} {\bibfnamefont {A.}~\bibnamefont {Teixeira}},\ }\href
  {\doibase 10.1088/1126-6708/2006/11/090} {\bibfield  {journal} {\bibinfo
  {journal} {JHEP}\ }\textbf {\bibinfo {volume} {0611}},\ \bibinfo {pages}
  {090} (\bibinfo {year} {2006})},\ \Eprint
  {http://arxiv.org/abs/hep-ph/0607263} {arXiv:hep-ph/0607263 [hep-ph]}
  \BibitemShut {NoStop}%
\bibitem [{\citenamefont {Calibbi}\ \emph {et~al.}(2007)\citenamefont
  {Calibbi}, \citenamefont {Faccia}, \citenamefont {Masiero},\ and\
  \citenamefont {Vempati}}]{Calibbi:2006ne}%
  \BibitemOpen
  \bibfield  {author} {\bibinfo {author} {\bibfnamefont {L.}~\bibnamefont
  {Calibbi}}, \bibinfo {author} {\bibfnamefont {A.}~\bibnamefont {Faccia}},
  \bibinfo {author} {\bibfnamefont {A.}~\bibnamefont {Masiero}}, \ and\
  \bibinfo {author} {\bibfnamefont {S.}~\bibnamefont {Vempati}},\ }\href
  {\doibase 10.1088/1126-6708/2007/07/012} {\bibfield  {journal} {\bibinfo
  {journal} {JHEP}\ }\textbf {\bibinfo {volume} {0707}},\ \bibinfo {pages}
  {012} (\bibinfo {year} {2007})},\ \Eprint
  {http://arxiv.org/abs/hep-ph/0610241} {arXiv:hep-ph/0610241 [hep-ph]}
  \BibitemShut {NoStop}%
\bibitem [{\citenamefont {Masiero}\ \emph {et~al.}(2007)\citenamefont
  {Masiero}, \citenamefont {Vempati},\ and\ \citenamefont {Vives}}]{reviews1}%
  \BibitemOpen
  \bibfield  {author} {\bibinfo {author} {\bibfnamefont {A.}~\bibnamefont
  {Masiero}}, \bibinfo {author} {\bibfnamefont {S.}~\bibnamefont {Vempati}}, \
  and\ \bibinfo {author} {\bibfnamefont {O.}~\bibnamefont {Vives}},\
  }\href@noop {} {\  (\bibinfo {year} {2007})},\ \Eprint
  {http://arxiv.org/abs/0711.2903} {arXiv:0711.2903 [hep-ph]} \BibitemShut
  {NoStop}%
\bibitem [{\citenamefont {Arganda}\ \emph {et~al.}(2007)\citenamefont
  {Arganda}, \citenamefont {Herrero},\ and\ \citenamefont
  {Teixeira}}]{herrero}%
  \BibitemOpen
  \bibfield  {author} {\bibinfo {author} {\bibfnamefont {E.}~\bibnamefont
  {Arganda}}, \bibinfo {author} {\bibfnamefont {M.}~\bibnamefont {Herrero}}, \
  and\ \bibinfo {author} {\bibfnamefont {A.}~\bibnamefont {Teixeira}},\ }\href
  {\doibase 10.1088/1126-6708/2007/10/104} {\bibfield  {journal} {\bibinfo
  {journal} {JHEP}\ }\textbf {\bibinfo {volume} {0710}},\ \bibinfo {pages}
  {104} (\bibinfo {year} {2007})},\ \Eprint {http://arxiv.org/abs/0707.2955}
  {arXiv:0707.2955 [hep-ph]} \BibitemShut {NoStop}%
\bibitem [{\citenamefont {Joshipura}\ \emph {et~al.}(2010)\citenamefont
  {Joshipura}, \citenamefont {Patel},\ and\ \citenamefont
  {Vempati}}]{Joshipura:2009gi}%
  \BibitemOpen
  \bibfield  {author} {\bibinfo {author} {\bibfnamefont {A.~S.}\ \bibnamefont
  {Joshipura}}, \bibinfo {author} {\bibfnamefont {K.~M.}\ \bibnamefont
  {Patel}}, \ and\ \bibinfo {author} {\bibfnamefont {S.~K.}\ \bibnamefont
  {Vempati}},\ }\href {\doibase 10.1016/j.physletb.2010.05.034} {\bibfield
  {journal} {\bibinfo  {journal} {Phys.Lett.}\ }\textbf {\bibinfo {volume}
  {B690}},\ \bibinfo {pages} {289} (\bibinfo {year} {2010})},\ \Eprint
  {http://arxiv.org/abs/0911.5618} {arXiv:0911.5618 [hep-ph]} \BibitemShut
  {NoStop}%
\bibitem [{\citenamefont {Biggio}\ and\ \citenamefont
  {Calibbi}(2010)}]{Biggio:2010me}%
  \BibitemOpen
  \bibfield  {author} {\bibinfo {author} {\bibfnamefont {C.}~\bibnamefont
  {Biggio}}\ and\ \bibinfo {author} {\bibfnamefont {L.}~\bibnamefont
  {Calibbi}},\ }\href {\doibase 10.1007/JHEP10(2010)037} {\bibfield  {journal}
  {\bibinfo  {journal} {JHEP}\ }\textbf {\bibinfo {volume} {1010}},\ \bibinfo
  {pages} {037} (\bibinfo {year} {2010})},\ \Eprint
  {http://arxiv.org/abs/1007.3750} {arXiv:1007.3750 [hep-ph]} \BibitemShut
  {NoStop}%
\bibitem [{\citenamefont {Esteves}\ \emph {et~al.}(2011)\citenamefont
  {Esteves}, \citenamefont {Romao}, \citenamefont {Hirsch}, \citenamefont
  {Staub},\ and\ \citenamefont {Porod}}]{Esteves:2010ff}%
  \BibitemOpen
  \bibfield  {author} {\bibinfo {author} {\bibfnamefont {J.}~\bibnamefont
  {Esteves}}, \bibinfo {author} {\bibfnamefont {J.}~\bibnamefont {Romao}},
  \bibinfo {author} {\bibfnamefont {M.}~\bibnamefont {Hirsch}}, \bibinfo
  {author} {\bibfnamefont {F.}~\bibnamefont {Staub}}, \ and\ \bibinfo {author}
  {\bibfnamefont {W.}~\bibnamefont {Porod}},\ }\href {\doibase
  10.1103/PhysRevD.83.013003} {\bibfield  {journal} {\bibinfo  {journal}
  {Phys.Rev.}\ }\textbf {\bibinfo {volume} {D83}},\ \bibinfo {pages} {013003}
  (\bibinfo {year} {2011})},\ \Eprint {http://arxiv.org/abs/1010.6000}
  {arXiv:1010.6000 [hep-ph]} \BibitemShut {NoStop}%
\bibitem [{\citenamefont {Cannoni}\ \emph {et~al.}(2013)\citenamefont
  {Cannoni}, \citenamefont {Ellis}, \citenamefont {Gomez},\ and\ \citenamefont
  {Lola}}]{Cannoni:2013gq}%
  \BibitemOpen
  \bibfield  {author} {\bibinfo {author} {\bibfnamefont {M.}~\bibnamefont
  {Cannoni}}, \bibinfo {author} {\bibfnamefont {J.}~\bibnamefont {Ellis}},
  \bibinfo {author} {\bibfnamefont {M.~E.}\ \bibnamefont {Gomez}}, \ and\
  \bibinfo {author} {\bibfnamefont {S.}~\bibnamefont {Lola}},\ }\href@noop {}
  {\  (\bibinfo {year} {2013})},\ \Eprint {http://arxiv.org/abs/1301.6002}
  {arXiv:1301.6002 [hep-ph]} \BibitemShut {NoStop}%
\bibitem [{\citenamefont {Calibbi}\ \emph {et~al.}(2012)\citenamefont
  {Calibbi}, \citenamefont {Chowdhury}, \citenamefont {Masiero}, \citenamefont
  {Patel},\ and\ \citenamefont {Vempati}}]{Calibbi:2012gr}%
  \BibitemOpen
  \bibfield  {author} {\bibinfo {author} {\bibfnamefont {L.}~\bibnamefont
  {Calibbi}}, \bibinfo {author} {\bibfnamefont {D.}~\bibnamefont {Chowdhury}},
  \bibinfo {author} {\bibfnamefont {A.}~\bibnamefont {Masiero}}, \bibinfo
  {author} {\bibfnamefont {K.}~\bibnamefont {Patel}}, \ and\ \bibinfo {author}
  {\bibfnamefont {S.}~\bibnamefont {Vempati}},\ }\href {\doibase
  10.1007/JHEP11(2012)040} {\bibfield  {journal} {\bibinfo  {journal} {JHEP}\
  }\textbf {\bibinfo {volume} {1211}},\ \bibinfo {pages} {040} (\bibinfo {year}
  {2012})},\ \Eprint {http://arxiv.org/abs/1207.7227} {arXiv:1207.7227
  [hep-ph]} \BibitemShut {NoStop}%
\bibitem [{\citenamefont {Rossi}(2002)}]{Rossi:2002zb}%
  \BibitemOpen
  \bibfield  {author} {\bibinfo {author} {\bibfnamefont {A.}~\bibnamefont
  {Rossi}},\ }\href {\doibase 10.1103/PhysRevD.66.075003} {\bibfield  {journal}
  {\bibinfo  {journal} {Phys.Rev.}\ }\textbf {\bibinfo {volume} {D66}},\
  \bibinfo {pages} {075003} (\bibinfo {year} {2002})},\ \Eprint
  {http://arxiv.org/abs/hep-ph/0207006} {arXiv:hep-ph/0207006 [hep-ph]}
  \BibitemShut {NoStop}%
\bibitem [{\citenamefont {Joaquim}\ and\ \citenamefont
  {Rossi}(2006)}]{Joaquim:2006uz}%
  \BibitemOpen
  \bibfield  {author} {\bibinfo {author} {\bibfnamefont {F.}~\bibnamefont
  {Joaquim}}\ and\ \bibinfo {author} {\bibfnamefont {A.}~\bibnamefont
  {Rossi}},\ }\href {\doibase 10.1103/PhysRevLett.97.181801} {\bibfield
  {journal} {\bibinfo  {journal} {Phys.Rev.Lett.}\ }\textbf {\bibinfo {volume}
  {97}},\ \bibinfo {pages} {181801} (\bibinfo {year} {2006})},\ \Eprint
  {http://arxiv.org/abs/hep-ph/0604083} {arXiv:hep-ph/0604083 [hep-ph]}
  \BibitemShut {NoStop}%
\bibitem [{\citenamefont {Joaquim}\ and\ \citenamefont
  {Rossi}(2007)}]{Joaquim:2006mn}%
  \BibitemOpen
  \bibfield  {author} {\bibinfo {author} {\bibfnamefont {F.}~\bibnamefont
  {Joaquim}}\ and\ \bibinfo {author} {\bibfnamefont {A.}~\bibnamefont
  {Rossi}},\ }\href {\doibase 10.1016/j.nuclphysb.2006.11.030} {\bibfield
  {journal} {\bibinfo  {journal} {Nucl.Phys.}\ }\textbf {\bibinfo {volume}
  {B765}},\ \bibinfo {pages} {71} (\bibinfo {year} {2007})},\ \Eprint
  {http://arxiv.org/abs/hep-ph/0607298} {arXiv:hep-ph/0607298 [hep-ph]}
  \BibitemShut {NoStop}%
\bibitem [{\citenamefont {Joaquim}(2010)}]{Joaquim:2009vp}%
  \BibitemOpen
  \bibfield  {author} {\bibinfo {author} {\bibfnamefont {F.}~\bibnamefont
  {Joaquim}},\ }\href {\doibase 10.1007/JHEP06(2010)079} {\bibfield  {journal}
  {\bibinfo  {journal} {JHEP}\ }\textbf {\bibinfo {volume} {1006}},\ \bibinfo
  {pages} {079} (\bibinfo {year} {2010})},\ \Eprint
  {http://arxiv.org/abs/0912.3427} {arXiv:0912.3427 [hep-ph]} \BibitemShut
  {NoStop}%
\bibitem [{\citenamefont {Arganda}\ and\ \citenamefont
  {Herrero}(2006)}]{Arganda:2005ji}%
  \BibitemOpen
  \bibfield  {author} {\bibinfo {author} {\bibfnamefont {E.}~\bibnamefont
  {Arganda}}\ and\ \bibinfo {author} {\bibfnamefont {M.~J.}\ \bibnamefont
  {Herrero}},\ }\href {\doibase 10.1103/PhysRevD.73.055003} {\bibfield
  {journal} {\bibinfo  {journal} {Phys.Rev.}\ }\textbf {\bibinfo {volume}
  {D73}},\ \bibinfo {pages} {055003} (\bibinfo {year} {2006})},\ \Eprint
  {http://arxiv.org/abs/hep-ph/0510405} {arXiv:hep-ph/0510405 [hep-ph]}
  \BibitemShut {NoStop}%
\bibitem [{\citenamefont {Hirsch}\ \emph {et~al.}(2008)\citenamefont {Hirsch},
  \citenamefont {Kaneko},\ and\ \citenamefont {Porod}}]{Hirsch:2008gh}%
  \BibitemOpen
  \bibfield  {author} {\bibinfo {author} {\bibfnamefont {M.}~\bibnamefont
  {Hirsch}}, \bibinfo {author} {\bibfnamefont {S.}~\bibnamefont {Kaneko}}, \
  and\ \bibinfo {author} {\bibfnamefont {W.}~\bibnamefont {Porod}},\ }\href
  {\doibase 10.1103/PhysRevD.78.093004} {\bibfield  {journal} {\bibinfo
  {journal} {Phys.Rev.}\ }\textbf {\bibinfo {volume} {D78}},\ \bibinfo {pages}
  {093004} (\bibinfo {year} {2008})},\ \Eprint {http://arxiv.org/abs/0806.3361}
  {arXiv:0806.3361 [hep-ph]} \BibitemShut {NoStop}%
\bibitem [{\citenamefont {Esteves}\ \emph {et~al.}(2009)\citenamefont
  {Esteves}, \citenamefont {Kaneko}, \citenamefont {Romao}, \citenamefont
  {Hirsch},\ and\ \citenamefont {Porod}}]{Esteves:2009qr}%
  \BibitemOpen
  \bibfield  {author} {\bibinfo {author} {\bibfnamefont {J.}~\bibnamefont
  {Esteves}}, \bibinfo {author} {\bibfnamefont {S.}~\bibnamefont {Kaneko}},
  \bibinfo {author} {\bibfnamefont {J.}~\bibnamefont {Romao}}, \bibinfo
  {author} {\bibfnamefont {M.}~\bibnamefont {Hirsch}}, \ and\ \bibinfo {author}
  {\bibfnamefont {W.}~\bibnamefont {Porod}},\ }\href {\doibase
  10.1103/PhysRevD.80.095003} {\bibfield  {journal} {\bibinfo  {journal}
  {Phys.Rev.}\ }\textbf {\bibinfo {volume} {D80}},\ \bibinfo {pages} {095003}
  (\bibinfo {year} {2009})},\ \Eprint {http://arxiv.org/abs/0907.5090}
  {arXiv:0907.5090 [hep-ph]} \BibitemShut {NoStop}%
\bibitem [{\citenamefont {Calibbi}\ \emph {et~al.}(2009)\citenamefont
  {Calibbi}, \citenamefont {Frigerio}, \citenamefont {Lavignac},\ and\
  \citenamefont {Romanino}}]{Calibbi:2009wk}%
  \BibitemOpen
  \bibfield  {author} {\bibinfo {author} {\bibfnamefont {L.}~\bibnamefont
  {Calibbi}}, \bibinfo {author} {\bibfnamefont {M.}~\bibnamefont {Frigerio}},
  \bibinfo {author} {\bibfnamefont {S.}~\bibnamefont {Lavignac}}, \ and\
  \bibinfo {author} {\bibfnamefont {A.}~\bibnamefont {Romanino}},\ }\href
  {\doibase 10.1088/1126-6708/2009/12/057} {\bibfield  {journal} {\bibinfo
  {journal} {JHEP}\ }\textbf {\bibinfo {volume} {0912}},\ \bibinfo {pages}
  {057} (\bibinfo {year} {2009})},\ \Eprint {http://arxiv.org/abs/0910.0377}
  {arXiv:0910.0377 [hep-ph]} \BibitemShut {NoStop}%
\bibitem [{\citenamefont {Hirsch}\ \emph {et~al.}(2012)\citenamefont {Hirsch},
  \citenamefont {Joaquim},\ and\ \citenamefont {Vicente}}]{Hirsch:2012ti}%
  \BibitemOpen
  \bibfield  {author} {\bibinfo {author} {\bibfnamefont {M.}~\bibnamefont
  {Hirsch}}, \bibinfo {author} {\bibfnamefont {F.}~\bibnamefont {Joaquim}}, \
  and\ \bibinfo {author} {\bibfnamefont {A.}~\bibnamefont {Vicente}},\ }\href
  {\doibase 10.1007/JHEP11(2012)105} {\bibfield  {journal} {\bibinfo  {journal}
  {JHEP}\ }\textbf {\bibinfo {volume} {1211}},\ \bibinfo {pages} {105}
  (\bibinfo {year} {2012})},\ \Eprint {http://arxiv.org/abs/1207.6635}
  {arXiv:1207.6635 [hep-ph]} \BibitemShut {NoStop}%
\bibitem [{\citenamefont {Abe}\ \emph {et~al.}(2011)\citenamefont {Abe} \emph
  {et~al.}}]{Abe:2011sj}%
  \BibitemOpen
  \bibfield  {author} {\bibinfo {author} {\bibfnamefont {K.}~\bibnamefont
  {Abe}} \emph {et~al.} (\bibinfo {collaboration} {T2K Collaboration}),\ }\href
  {\doibase 10.1103/PhysRevLett.107.041801} {\bibfield  {journal} {\bibinfo
  {journal} {Phys.Rev.Lett.}\ }\textbf {\bibinfo {volume} {107}},\ \bibinfo
  {pages} {041801} (\bibinfo {year} {2011})},\ \Eprint
  {http://arxiv.org/abs/1106.2822} {arXiv:1106.2822 [hep-ex]} \BibitemShut
  {NoStop}%
\bibitem [{\citenamefont {Adamson}\ \emph {et~al.}(2011)\citenamefont {Adamson}
  \emph {et~al.}}]{Adamson:2011qu}%
  \BibitemOpen
  \bibfield  {author} {\bibinfo {author} {\bibfnamefont {P.}~\bibnamefont
  {Adamson}} \emph {et~al.} (\bibinfo {collaboration} {MINOS Collaboration}),\
  }\href {\doibase 10.1103/PhysRevLett.107.181802} {\bibfield  {journal}
  {\bibinfo  {journal} {Phys.Rev.Lett.}\ }\textbf {\bibinfo {volume} {107}},\
  \bibinfo {pages} {181802} (\bibinfo {year} {2011})},\ \Eprint
  {http://arxiv.org/abs/1108.0015} {arXiv:1108.0015 [hep-ex]} \BibitemShut
  {NoStop}%
\bibitem [{\citenamefont {Abe}\ \emph {et~al.}(2012)\citenamefont {Abe} \emph
  {et~al.}}]{Abe:2011fz}%
  \BibitemOpen
  \bibfield  {author} {\bibinfo {author} {\bibfnamefont {Y.}~\bibnamefont
  {Abe}} \emph {et~al.} (\bibinfo {collaboration} {DOUBLE-CHOOZ
  Collaboration}),\ }\href {\doibase 10.1103/PhysRevLett.108.131801} {\bibfield
   {journal} {\bibinfo  {journal} {Phys.Rev.Lett.}\ }\textbf {\bibinfo {volume}
  {108}},\ \bibinfo {pages} {131801} (\bibinfo {year} {2012})},\ \Eprint
  {http://arxiv.org/abs/1112.6353} {arXiv:1112.6353 [hep-ex]} \BibitemShut
  {NoStop}%
\bibitem [{\citenamefont {An}\ \emph {et~al.}(2012)\citenamefont {An} \emph
  {et~al.}}]{An:2012eh}%
  \BibitemOpen
  \bibfield  {author} {\bibinfo {author} {\bibfnamefont {F.}~\bibnamefont {An}}
  \emph {et~al.} (\bibinfo {collaboration} {DAYA-BAY Collaboration}),\ }\href
  {\doibase 10.1103/PhysRevLett.108.171803} {\bibfield  {journal} {\bibinfo
  {journal} {Phys.Rev.Lett.}\ }\textbf {\bibinfo {volume} {108}},\ \bibinfo
  {pages} {171803} (\bibinfo {year} {2012})},\ \Eprint
  {http://arxiv.org/abs/1203.1669} {arXiv:1203.1669 [hep-ex]} \BibitemShut
  {NoStop}%
\bibitem [{\citenamefont {Ahn}\ \emph {et~al.}(2012)\citenamefont {Ahn} \emph
  {et~al.}}]{Ahn:2012nd}%
  \BibitemOpen
  \bibfield  {author} {\bibinfo {author} {\bibfnamefont {J.}~\bibnamefont
  {Ahn}} \emph {et~al.} (\bibinfo {collaboration} {RENO collaboration}),\
  }\href {\doibase 10.1103/PhysRevLett.108.191802} {\bibfield  {journal}
  {\bibinfo  {journal} {Phys.Rev.Lett.}\ }\textbf {\bibinfo {volume} {108}},\
  \bibinfo {pages} {191802} (\bibinfo {year} {2012})},\ \Eprint
  {http://arxiv.org/abs/1204.0626} {arXiv:1204.0626 [hep-ex]} \BibitemShut
  {NoStop}%
\bibitem [{\citenamefont {Gonzalez-Garcia}\ \emph {et~al.}(2012)\citenamefont
  {Gonzalez-Garcia}, \citenamefont {Maltoni}, \citenamefont {Salvado},\ and\
  \citenamefont {Schwetz}}]{GonzalezGarcia:2012sz}%
  \BibitemOpen
  \bibfield  {author} {\bibinfo {author} {\bibfnamefont {M.}~\bibnamefont
  {Gonzalez-Garcia}}, \bibinfo {author} {\bibfnamefont {M.}~\bibnamefont
  {Maltoni}}, \bibinfo {author} {\bibfnamefont {J.}~\bibnamefont {Salvado}}, \
  and\ \bibinfo {author} {\bibfnamefont {T.}~\bibnamefont {Schwetz}},\ }\href
  {\doibase 10.1007/JHEP12(2012)123} {\bibfield  {journal} {\bibinfo  {journal}
  {JHEP}\ }\textbf {\bibinfo {volume} {1212}},\ \bibinfo {pages} {123}
  (\bibinfo {year} {2012})},\ \Eprint {http://arxiv.org/abs/1209.3023}
  {arXiv:1209.3023 [hep-ph]} \BibitemShut {NoStop}%
\bibitem [{\citenamefont {Adam}\ \emph {et~al.}(2013)\citenamefont {Adam} \emph
  {et~al.}}]{Adam:2013mnn}%
  \BibitemOpen
  \bibfield  {author} {\bibinfo {author} {\bibfnamefont {J.}~\bibnamefont
  {Adam}} \emph {et~al.} (\bibinfo {collaboration} {MEG Collaboration}),\
  }\href@noop {} {\  (\bibinfo {year} {2013})},\ \Eprint
  {http://arxiv.org/abs/1303.0754} {arXiv:1303.0754 [hep-ex]} \BibitemShut
  {NoStop}%
\bibitem [{\citenamefont {Baldini}\ \emph {et~al.}(2013)\citenamefont
  {Baldini}, \citenamefont {Cei}, \citenamefont {Cerri}, \citenamefont
  {Dussoni}, \citenamefont {Galli} \emph {et~al.}}]{Baldini:2013ke}%
  \BibitemOpen
  \bibfield  {author} {\bibinfo {author} {\bibfnamefont {A.}~\bibnamefont
  {Baldini}}, \bibinfo {author} {\bibfnamefont {F.}~\bibnamefont {Cei}},
  \bibinfo {author} {\bibfnamefont {C.}~\bibnamefont {Cerri}}, \bibinfo
  {author} {\bibfnamefont {S.}~\bibnamefont {Dussoni}}, \bibinfo {author}
  {\bibfnamefont {L.}~\bibnamefont {Galli}},  \emph {et~al.},\ }\href@noop {}
  {\  (\bibinfo {year} {2013})},\ \Eprint {http://arxiv.org/abs/1301.7225}
  {arXiv:1301.7225 [physics.ins-det]} \BibitemShut {NoStop}%
\bibitem [{\citenamefont {Amhis}\ \emph {et~al.}(2012)\citenamefont {Amhis}
  \emph {et~al.}}]{Amhis:2012bh}%
  \BibitemOpen
  \bibfield  {author} {\bibinfo {author} {\bibfnamefont {Y.}~\bibnamefont
  {Amhis}} \emph {et~al.} (\bibinfo {collaboration} {Heavy Flavor Averaging
  Group}),\ }\href@noop {} {\  (\bibinfo {year} {2012})},\ \Eprint
  {http://arxiv.org/abs/1207.1158} {arXiv:1207.1158 [hep-ex]} \BibitemShut
  {NoStop}%
\bibitem [{\citenamefont {Brodzicka}\ \emph {et~al.}(2012)\citenamefont
  {Brodzicka} \emph {et~al.}}]{Brodzicka:2012jm}%
  \BibitemOpen
  \bibfield  {author} {\bibinfo {author} {\bibfnamefont {J.}~\bibnamefont
  {Brodzicka}} \emph {et~al.} (\bibinfo {collaboration} {Belle
  Collaboration}),\ }\href {\doibase 10.1093/ptep/pts072} {\bibfield  {journal}
  {\bibinfo  {journal} {PTEP}\ }\textbf {\bibinfo {volume} {2012}},\ \bibinfo
  {pages} {04D001} (\bibinfo {year} {2012})},\ \Eprint
  {http://arxiv.org/abs/1212.5342} {arXiv:1212.5342 [hep-ex]} \BibitemShut
  {NoStop}%
\bibitem [{\citenamefont {Bellgardt}\ \emph {et~al.}(1988)\citenamefont
  {Bellgardt} \emph {et~al.}}]{Bellgardt:1987du}%
  \BibitemOpen
  \bibfield  {author} {\bibinfo {author} {\bibfnamefont {U.}~\bibnamefont
  {Bellgardt}} \emph {et~al.} (\bibinfo {collaboration} {SINDRUM
  Collaboration}),\ }\href {\doibase 10.1016/0550-3213(88)90462-2} {\bibfield
  {journal} {\bibinfo  {journal} {Nucl.Phys.}\ }\textbf {\bibinfo {volume}
  {B299}},\ \bibinfo {pages} {1} (\bibinfo {year} {1988})}\BibitemShut
  {NoStop}%
\bibitem [{\citenamefont {Blondel}\ \emph {et~al.}(2013)\citenamefont
  {Blondel}, \citenamefont {Bravar}, \citenamefont {Pohl}, \citenamefont
  {Bachmann}, \citenamefont {Berger} \emph {et~al.}}]{Blondel:2013ia}%
  \BibitemOpen
  \bibfield  {author} {\bibinfo {author} {\bibfnamefont {A.}~\bibnamefont
  {Blondel}}, \bibinfo {author} {\bibfnamefont {A.}~\bibnamefont {Bravar}},
  \bibinfo {author} {\bibfnamefont {M.}~\bibnamefont {Pohl}}, \bibinfo {author}
  {\bibfnamefont {S.}~\bibnamefont {Bachmann}}, \bibinfo {author}
  {\bibfnamefont {N.}~\bibnamefont {Berger}},  \emph {et~al.},\ }\href@noop {}
  {\  (\bibinfo {year} {2013})},\ \Eprint {http://arxiv.org/abs/1301.6113}
  {arXiv:1301.6113 [physics.ins-det]} \BibitemShut {NoStop}%
\bibitem [{\citenamefont {Beringer}\ \emph {et~al.}(2012)\citenamefont
  {Beringer} \emph {et~al.}}]{Beringer:1900zz}%
  \BibitemOpen
  \bibfield  {author} {\bibinfo {author} {\bibfnamefont {J.}~\bibnamefont
  {Beringer}} \emph {et~al.} (\bibinfo {collaboration} {Particle Data Group}),\
  }\href {\doibase 10.1103/PhysRevD.86.010001} {\bibfield  {journal} {\bibinfo
  {journal} {Phys.Rev.}\ }\textbf {\bibinfo {volume} {D86}},\ \bibinfo {pages}
  {010001} (\bibinfo {year} {2012})}\BibitemShut {NoStop}%
\bibitem [{\citenamefont {Antusch}\ \emph {et~al.}(2003)\citenamefont
  {Antusch}, \citenamefont {Kersten}, \citenamefont {Lindner},\ and\
  \citenamefont {Ratz}}]{Antusch:2003kp}%
  \BibitemOpen
  \bibfield  {author} {\bibinfo {author} {\bibfnamefont {S.}~\bibnamefont
  {Antusch}}, \bibinfo {author} {\bibfnamefont {J.}~\bibnamefont {Kersten}},
  \bibinfo {author} {\bibfnamefont {M.}~\bibnamefont {Lindner}}, \ and\
  \bibinfo {author} {\bibfnamefont {M.}~\bibnamefont {Ratz}},\ }\href {\doibase
  10.1016/j.nuclphysb.2003.09.050} {\bibfield  {journal} {\bibinfo  {journal}
  {Nucl.Phys.}\ }\textbf {\bibinfo {volume} {B674}},\ \bibinfo {pages} {401}
  (\bibinfo {year} {2003})},\ \Eprint {http://arxiv.org/abs/hep-ph/0305273}
  {arXiv:hep-ph/0305273 [hep-ph]} \BibitemShut {NoStop}%
\bibitem [{\citenamefont {Porod}\ and\ \citenamefont
  {Staub}(2012)}]{Porod:2011nf}%
  \BibitemOpen
  \bibfield  {author} {\bibinfo {author} {\bibfnamefont {W.}~\bibnamefont
  {Porod}}\ and\ \bibinfo {author} {\bibfnamefont {F.}~\bibnamefont {Staub}},\
  }\href {\doibase 10.1016/j.cpc.2012.05.021} {\bibfield  {journal} {\bibinfo
  {journal} {Comput.Phys.Commun.}\ }\textbf {\bibinfo {volume} {183}},\
  \bibinfo {pages} {2458} (\bibinfo {year} {2012})},\ \Eprint
  {http://arxiv.org/abs/1104.1573} {arXiv:1104.1573 [hep-ph]} \BibitemShut
  {NoStop}%
\bibitem [{\citenamefont {Pierce}\ \emph {et~al.}(1997)\citenamefont {Pierce},
  \citenamefont {Bagger}, \citenamefont {Matchev},\ and\ \citenamefont
  {Zhang}}]{Pierce:1996zz}%
  \BibitemOpen
  \bibfield  {author} {\bibinfo {author} {\bibfnamefont {D.~M.}\ \bibnamefont
  {Pierce}}, \bibinfo {author} {\bibfnamefont {J.~A.}\ \bibnamefont {Bagger}},
  \bibinfo {author} {\bibfnamefont {K.~T.}\ \bibnamefont {Matchev}}, \ and\
  \bibinfo {author} {\bibfnamefont {R.-j.}\ \bibnamefont {Zhang}},\ }\href
  {\doibase 10.1016/S0550-3213(96)00683-9} {\bibfield  {journal} {\bibinfo
  {journal} {Nucl.Phys.}\ }\textbf {\bibinfo {volume} {B491}},\ \bibinfo
  {pages} {3} (\bibinfo {year} {1997})},\ \Eprint
  {http://arxiv.org/abs/hep-ph/9606211} {arXiv:hep-ph/9606211 [hep-ph]}
  \BibitemShut {NoStop}%
\bibitem [{\citenamefont {Dedes}\ and\ \citenamefont
  {Slavich}(2003)}]{Dedes:2002dy}%
  \BibitemOpen
  \bibfield  {author} {\bibinfo {author} {\bibfnamefont {A.}~\bibnamefont
  {Dedes}}\ and\ \bibinfo {author} {\bibfnamefont {P.}~\bibnamefont
  {Slavich}},\ }\href {\doibase 10.1016/S0550-3213(03)00173-1} {\bibfield
  {journal} {\bibinfo  {journal} {Nucl.Phys.}\ }\textbf {\bibinfo {volume}
  {B657}},\ \bibinfo {pages} {333} (\bibinfo {year} {2003})},\ \Eprint
  {http://arxiv.org/abs/hep-ph/0212132} {arXiv:hep-ph/0212132 [hep-ph]}
  \BibitemShut {NoStop}%
\bibitem [{\citenamefont {Brignole}\ \emph {et~al.}(2002)\citenamefont
  {Brignole}, \citenamefont {Degrassi}, \citenamefont {Slavich},\ and\
  \citenamefont {Zwirner}}]{Brignole:2001jy}%
  \BibitemOpen
  \bibfield  {author} {\bibinfo {author} {\bibfnamefont {A.}~\bibnamefont
  {Brignole}}, \bibinfo {author} {\bibfnamefont {G.}~\bibnamefont {Degrassi}},
  \bibinfo {author} {\bibfnamefont {P.}~\bibnamefont {Slavich}}, \ and\
  \bibinfo {author} {\bibfnamefont {F.}~\bibnamefont {Zwirner}},\ }\href
  {\doibase 10.1016/S0550-3213(02)00184-0} {\bibfield  {journal} {\bibinfo
  {journal} {Nucl.Phys.}\ }\textbf {\bibinfo {volume} {B631}},\ \bibinfo
  {pages} {195} (\bibinfo {year} {2002})},\ \Eprint
  {http://arxiv.org/abs/hep-ph/0112177} {arXiv:hep-ph/0112177 [hep-ph]}
  \BibitemShut {NoStop}%
\bibitem [{\citenamefont {Degrassi}\ \emph {et~al.}(2001)\citenamefont
  {Degrassi}, \citenamefont {Slavich},\ and\ \citenamefont
  {Zwirner}}]{Degrassi:2001yf}%
  \BibitemOpen
  \bibfield  {author} {\bibinfo {author} {\bibfnamefont {G.}~\bibnamefont
  {Degrassi}}, \bibinfo {author} {\bibfnamefont {P.}~\bibnamefont {Slavich}}, \
  and\ \bibinfo {author} {\bibfnamefont {F.}~\bibnamefont {Zwirner}},\ }\href
  {\doibase 10.1016/S0550-3213(01)00343-1} {\bibfield  {journal} {\bibinfo
  {journal} {Nucl.Phys.}\ }\textbf {\bibinfo {volume} {B611}},\ \bibinfo
  {pages} {403} (\bibinfo {year} {2001})},\ \Eprint
  {http://arxiv.org/abs/hep-ph/0105096} {arXiv:hep-ph/0105096 [hep-ph]}
  \BibitemShut {NoStop}%
\bibitem [{\citenamefont {Dedes}\ \emph {et~al.}(2003)\citenamefont {Dedes},
  \citenamefont {Degrassi},\ and\ \citenamefont {Slavich}}]{slavich}%
  \BibitemOpen
  \bibfield  {author} {\bibinfo {author} {\bibfnamefont {A.}~\bibnamefont
  {Dedes}}, \bibinfo {author} {\bibfnamefont {G.}~\bibnamefont {Degrassi}}, \
  and\ \bibinfo {author} {\bibfnamefont {P.}~\bibnamefont {Slavich}},\ }\href
  {\doibase 10.1016/j.nuclphysb.2003.08.033} {\bibfield  {journal} {\bibinfo
  {journal} {Nucl.Phys.}\ }\textbf {\bibinfo {volume} {B672}},\ \bibinfo
  {pages} {144} (\bibinfo {year} {2003})},\ \Eprint
  {http://arxiv.org/abs/hep-ph/0305127} {arXiv:hep-ph/0305127 [hep-ph]}
  \BibitemShut {NoStop}%
\bibitem [{\citenamefont {Aad}\ \emph {et~al.}(2012)\citenamefont {Aad} \emph
  {et~al.}}]{atlashiggs}%
  \BibitemOpen
  \bibfield  {author} {\bibinfo {author} {\bibfnamefont {G.}~\bibnamefont
  {Aad}} \emph {et~al.} (\bibinfo {collaboration} {ATLAS Collaboration}),\
  }\href {\doibase 10.1016/j.physletb.2012.08.020} {\bibfield  {journal}
  {\bibinfo  {journal} {Phys.Lett.}\ }\textbf {\bibinfo {volume} {B716}},\
  \bibinfo {pages} {1} (\bibinfo {year} {2012})},\ \Eprint
  {http://arxiv.org/abs/1207.7214} {arXiv:1207.7214 [hep-ex]} \BibitemShut
  {NoStop}%
\bibitem [{\citenamefont {Chatrchyan}\ \emph {et~al.}(2012)\citenamefont
  {Chatrchyan} \emph {et~al.}}]{cmshiggs}%
  \BibitemOpen
  \bibfield  {author} {\bibinfo {author} {\bibfnamefont {S.}~\bibnamefont
  {Chatrchyan}} \emph {et~al.} (\bibinfo {collaboration} {CMS Collaboration}),\
  }\href {\doibase 10.1016/j.physletb.2012.08.021} {\bibfield  {journal}
  {\bibinfo  {journal} {Phys.Lett.}\ }\textbf {\bibinfo {volume} {B716}},\
  \bibinfo {pages} {30} (\bibinfo {year} {2012})},\ \Eprint
  {http://arxiv.org/abs/1207.7235} {arXiv:1207.7235 [hep-ex]} \BibitemShut
  {NoStop}%
\bibitem [{\citenamefont {Arbey}\ \emph {et~al.}(2012)\citenamefont {Arbey},
  \citenamefont {Battaglia}, \citenamefont {Djouadi},\ and\ \citenamefont
  {Mahmoudi}}]{Arbey:2012dq}%
  \BibitemOpen
  \bibfield  {author} {\bibinfo {author} {\bibfnamefont {A.}~\bibnamefont
  {Arbey}}, \bibinfo {author} {\bibfnamefont {M.}~\bibnamefont {Battaglia}},
  \bibinfo {author} {\bibfnamefont {A.}~\bibnamefont {Djouadi}}, \ and\
  \bibinfo {author} {\bibfnamefont {F.}~\bibnamefont {Mahmoudi}},\ }\href
  {\doibase 10.1007/JHEP09(2012)107} {\bibfield  {journal} {\bibinfo  {journal}
  {JHEP}\ }\textbf {\bibinfo {volume} {1209}},\ \bibinfo {pages} {107}
  (\bibinfo {year} {2012})},\ \Eprint {http://arxiv.org/abs/1207.1348}
  {arXiv:1207.1348 [hep-ph]} \BibitemShut {NoStop}%
\bibitem [{\citenamefont {Adachi}\ \emph {et~al.}(2012)\citenamefont {Adachi}
  \emph {et~al.}}]{Adachi:2012mm}%
  \BibitemOpen
  \bibfield  {author} {\bibinfo {author} {\bibfnamefont {I.}~\bibnamefont
  {Adachi}} \emph {et~al.} (\bibinfo {collaboration} {Belle Collaboration}),\
  }\href@noop {} {\  (\bibinfo {year} {2012})},\ \Eprint
  {http://arxiv.org/abs/1208.4678} {arXiv:1208.4678 [hep-ex]} \BibitemShut
  {NoStop}%
\bibitem [{\citenamefont {Yook}()}]{Btaunu}%
  \BibitemOpen
  \bibfield  {author} {\bibinfo {author} {\bibfnamefont {Y.}~\bibnamefont
  {Yook}} (\bibinfo {collaboration} {Belle Collaboration}),\ }\href@noop {}
  {\bibinfo  {journal} {talk given at ICHEP 2012}\ }\BibitemShut {NoStop}%
\bibitem [{\citenamefont {Aaij}\ \emph {et~al.}(2013)\citenamefont {Aaij} \emph
  {et~al.}}]{Aaij:2012nna}%
  \BibitemOpen
\bibfield  {journal} {  }\bibfield  {author} {\bibinfo {author} {\bibfnamefont
  {R.}~\bibnamefont {Aaij}} \emph {et~al.} (\bibinfo {collaboration} {LHCb
  Collaboration}),\ }\href {\doibase 10.1103/PhysRevLett.110.021801} {\bibfield
   {journal} {\bibinfo  {journal} {Phys.Rev.Lett.}\ }\textbf {\bibinfo {volume}
  {110}},\ \bibinfo {pages} {021801} (\bibinfo {year} {2013})},\ \Eprint
  {http://arxiv.org/abs/1211.2674} {arXiv:1211.2674 [hep-ex]} \BibitemShut
  {NoStop}%
\end{thebibliography}%

\end{document}